\input epsf.sty
\epsfverbosetrue
\ifx\mnmacrosloaded\undefined 
%
%
%
%

\catcode `\@=11 

\def\@version{1.4}
\def\@verdate{22nd Feb 1994}

%
%
%
%


\newif\ifprod@font

\ifx\@typeface\undefined
  \def\@typeface{Comp. Modern}\prod@fontfalse
\else
  \prod@fonttrue 
\fi

\def\newfam{\alloc@8\fam\chardef\sixt@@n} 

\ifprod@font
\font\fiverm=mtr10 at 5pt
\font\fivebf=mtbx10 at 5pt
\font\fiveit=mtti10 at 5pt
\font\fivesl=mtsl10 at 5pt
\font\fivett=mttt10 at 5pt     \hyphenchar\fivett=-1
\font\fivecsc=mtcsc10 at 5pt
\font\fivesf=mtss10 at 5pt
\font\fivei=mtmi10 at 5pt      \skewchar\fivei='177
\font\fivemib=mtmib10 at 5pt   \skewchar\fivemib='177
\font\fivesy=mtsy10 at 5pt     \skewchar\fivesy='60
\font\fivesyb=mtbsy10 at 5pt   \skewchar\fivesyb='60

\font\sixrm=mtr10 at 6pt
\font\sixbf=mtbx10 at 6pt
\font\sixit=mtti10 at 6pt
\font\sixsl=mtsl10 at 6pt
\font\sixtt=mttt10 at 6pt      \hyphenchar\sixtt=-1
\font\sixcsc=mtcsc10 at 6pt
\font\sixsf=mtss10 at 6pt
\font\sixi=mtmi10 at 6pt       \skewchar\sixi='177
\font\sixmib=mtmib10 at 6pt    \skewchar\sixmib='177
\font\sixsy=mtsy10 at 6pt      \skewchar\sixsy='60
\font\sixsyb=mtbsy10 at 6pt    \skewchar\sixsyb='60

\font\sevenrm=mtr10 at 7pt
\font\sevenbf=mtbx10 at 7pt
\font\sevenit=mtti10 at 7pt
\font\sevensl=mtsl10 at 7pt
\font\seventt=mttt10 at 7pt     \hyphenchar\seventt=-1
\font\sevencsc=mtcsc10 at 7pt
\font\sevensf=mtss10 at 7pt
\font\seveni=mtmi10 at 7pt      \skewchar\seveni='177
\font\sevenmib=mtmib10 at 7pt   \skewchar\sevenmib='177
\font\sevensy=mtsy10 at 7pt     \skewchar\sevensy='60
\font\sevensyb=mtbsy10 at 7pt   \skewchar\sevensyb='60

\font\eightrm=mtr10 at 8pt
\font\eightbf=mtbx10 at 8pt
\font\eightit=mtti10 at 8pt
\font\eighti=mtmi10 at 8pt      \skewchar\eighti='177
\font\eightmib=mtmib10 at 8pt   \skewchar\eightmib='177
\font\eightsy=mtsy10 at 8pt     \skewchar\eightsy='60
\font\eightsyb=mtbsy10 at 8pt   \skewchar\eightsyb='60
\font\eightsl=mtsl10 at 8pt
\font\eighttt=mttt10 at 8pt     \hyphenchar\eighttt=-1
\font\eightcsc=mtcsc10 at 8pt
\font\eightsf=mtss10 at 8pt

\font\ninerm=mtr10 at 9pt
\font\ninebf=mtbx10 at 9pt
\font\nineit=mtti10 at 9pt
\font\ninei=mtmi10 at 9pt      \skewchar\ninei='177
\font\ninemib=mtmib10 at 9pt   \skewchar\ninemib='177
\font\ninesy=mtsy10 at 9pt     \skewchar\ninesy='60
\font\ninesyb=mtbsy10 at 9pt   \skewchar\ninesyb='60
\font\ninesl=mtsl10 at 9pt
\font\ninett=mttt10 at 9pt     \hyphenchar\ninett=-1
\font\ninecsc=mtcsc10 at 9pt
\font\ninesf=mtss10 at 9pt

\font\tenrm=mtr10
\font\tenbf=mtbx10
\font\tenit=mtti10
\font\teni=mtmi10		\skewchar\teni='177
\font\tenmib=mtmib10	\skewchar\tenmib='177
\font\tensy=mtsy10		\skewchar\tensy='60
\font\tensyb=mtbsy10	\skewchar\tensyb='60
\font\tenex=cmex10
\font\tensl=mtsl10
\font\tentt=mttt10		\hyphenchar\tentt=-1
\font\tencsc=mtcsc10
\font\tensf=mtss10

\font\elevenrm=mtr10 at 11pt
\font\elevenbf=mtbx10 at 11pt
\font\elevenit=mtti10 at 11pt
\font\eleveni=mtmi10 at 11pt      \skewchar\eleveni='177
\font\elevenmib=mtmib10 at 11pt   \skewchar\elevenmib='177
\font\elevensy=mtsy10 at 11pt     \skewchar\elevensy='60
\font\elevensyb=mtbsy10 at 11pt   \skewchar\elevensyb='60
\font\elevensl=mtsl10 at 11pt
\font\eleventt=mttt10 at 11pt     \hyphenchar\eleventt=-1
\font\elevencsc=mtcsc10 at 11pt
\font\elevensf=mtss10 at 11pt

\font\twelverm=mtr10 at 12pt
\font\twelvebf=mtbx10 at 12pt
\font\twelveit=mtti10 at 12pt
\font\twelvesl=mtsl10 at 12pt
\font\twelvett=mttt10 at 12pt     \hyphenchar\twelvett=-1
\font\twelvecsc=mtcsc10 at 12pt
\font\twelvesf=mtss10 at 12pt
\font\twelvei=mtmi10 at 12pt      \skewchar\twelvei='177
\font\twelvemib=mtmib10 at 12pt   \skewchar\twelvemib='177
\font\twelvesy=mtsy10 at 12pt     \skewchar\twelvesy='60
\font\twelvesyb=mtbsy10 at 12pt   \skewchar\twelvesyb='60

\font\fourteenrm=mtr10 at 14pt
\font\fourteenbf=mtbx10 at 14pt
\font\fourteenit=mtti10 at 14pt
\font\fourteeni=mtmi10 at 14pt      \skewchar\fourteeni='177
\font\fourteenmib=mtmib10 at 14pt   \skewchar\fourteenmib='177
\font\fourteensy=mtsy10 at 14pt     \skewchar\fourteensy='60
\font\fourteensyb=mtbsy10 at 14pt   \skewchar\fourteensyb='60
\font\fourteensl=mtsl10 at 14pt
\font\fourteentt=mttt10 at 14pt     \hyphenchar\fourteentt=-1
\font\fourteencsc=mtcsc10 at 14pt
\font\fourteensf=mtss10 at 14pt

\font\seventeenrm=mtr10 at 17pt
\font\seventeenbf=mtbx10 at 17pt
\font\seventeenit=mtti10 at 17pt
\font\seventeeni=mtmi10 at 17pt      \skewchar\seventeeni='177
\font\seventeenmib=mtmib10 at 17pt   \skewchar\seventeenmib='177
\font\seventeensy=mtsy10 at 17pt     \skewchar\seventeensy='60
\font\seventeensyb=mtbsy10 at 17pt   \skewchar\seventeensyb='60
\font\seventeensl=mtsl10 at 17pt
\font\seventeentt=mttt10 at 17pt     \hyphenchar\seventeentt=-1
\font\seventeencsc=mtcsc10 at 17pt
\font\seventeensf=mtss10 at 17pt


\newfam\xmfam
\newfam\ymfam

\font\fivexm=mtxm10 at 5pt
\font\sixxm=mtxm10 at 6pt
\font\sevenxm=mtxm10 at 7pt
\font\eightxm=mtxm10 at 8pt
\font\ninexm=mtxm10 at 9pt
\font\tenxm=mtxm10
\font\elevenxm=mtxm10 at 11pt
\font\twelvexm=mtxm10 at 12pt
\font\fourteenxm=mtxm10 at 14pt
\font\seventeenxm=mtxm10 at 17pt

\font\fiveym=mtym10 at 5pt
\font\sixym=mtym10 at 6pt
\font\sevenym=mtym10 at 7pt
\font\eightym=mtym10 at 8pt
\font\nineym=mtym10 at 9pt
\font\tenym=mtym10
\font\elevenym=mtym10 at 11pt
\font\twelveym=mtym10 at 12pt
\font\fourteenym=mtym10 at 14pt
\font\seventeenym=mtym10 at 17pt
\else
\font\fiverm=cmr5
\font\fivei=cmmi5             \skewchar\fivei='177
\font\fivemib=cmmib10 at 5pt  \skewchar\fivemib='177
\font\fivesy=cmsy5            \skewchar\fivesy='60
\font\fivesyb=cmbsy10 at 5pt  \skewchar\fivesyb='60
\font\fivebf=cmbx5

\font\sixrm=cmr6
\font\sixi=cmmi6             \skewchar\sixi='177
\font\sixmib=cmmib10 at 6pt  \skewchar\sixmib='177
\font\sixsy=cmsy6            \skewchar\sixsy='60
\font\sixsyb=cmbsy10 at 6pt  \skewchar\sixsyb='60
\font\sixbf=cmbx6

\font\sevenrm=cmr7
\font\seveni=cmmi7             \skewchar\seveni='177
\font\sevenmib=cmmib10 at 7pt  \skewchar\sevenmib='177
\font\sevensy=cmsy7            \skewchar\sevensy='60
\font\sevensyb=cmbsy10 at 7pt  \skewchar\sevensyb='60
\font\sevenbf=cmbx7

\font\eightrm=cmr8
\font\eightbf=cmbx8
\font\eightit=cmti8
\font\eighti=cmmi8			\skewchar\eighti='177
\font\eightmib=cmmib10 at 8pt	\skewchar\eightmib='177
\font\eightsy=cmsy8			\skewchar\eightsy='60
\font\eightsyb=cmbsy10 at 8pt	\skewchar\eightsyb='60
\font\eightsl=cmsl8
\font\eighttt=cmtt8			\hyphenchar\eighttt=-1
\font\eightcsc=cmcsc10 at 8pt
\font\eightsf=cmss8

\font\ninerm=cmr9
\font\ninebf=cmbx9
\font\nineit=cmti9
\font\ninei=cmmi9			\skewchar\ninei='177
\font\ninemib=cmmib10 at 9pt	\skewchar\ninemib='177
\font\ninesy=cmsy9			\skewchar\ninesy='60
\font\ninesyb=cmbsy10 at 9pt	\skewchar\ninesyb='60
\font\ninesl=cmsl9
\font\ninett=cmtt9			\hyphenchar\ninett=-1
\font\ninecsc=cmcsc10 at 9pt
\font\ninesf=cmss9

\font\tenrm=cmr10
\font\tenbf=cmbx10
\font\tenit=cmti10
\font\teni=cmmi10		\skewchar\teni='177
\font\tenmib=cmmib10	\skewchar\tenmib='177
\font\tensy=cmsy10		\skewchar\tensy='60
\font\tensyb=cmbsy10	\skewchar\tensyb='60
\font\tenex=cmex10
\font\tensl=cmsl10
\font\tentt=cmtt10		\hyphenchar\tentt=-1
\font\tencsc=cmcsc10
\font\tensf=cmss10

\font\elevenrm=cmr10 scaled \magstephalf
\font\elevenbf=cmbx10 scaled \magstephalf
\font\elevenit=cmti10 scaled \magstephalf
\font\eleveni=cmmi10 scaled \magstephalf	\skewchar\eleveni='177
\font\elevenmib=cmmib10 scaled \magstephalf	\skewchar\elevenmib='177
\font\elevensy=cmsy10 scaled \magstephalf	\skewchar\elevensy='60
\font\elevensyb=cmbsy10 scaled \magstephalf	\skewchar\elevensyb='60
\font\elevensl=cmsl10 scaled \magstephalf
\font\eleventt=cmtt10 scaled \magstephalf	\hyphenchar\eleventt=-1
\font\elevencsc=cmcsc10 scaled \magstephalf
\font\elevensf=cmss10 scaled \magstephalf

\font\twelverm=cmr10 scaled \magstep1
\font\twelvebf=cmbx10 scaled \magstep1
\font\twelvei=cmmi10 scaled \magstep1      \skewchar\twelvei='177
\font\twelvemib=cmmib10 scaled \magstep1   \skewchar\twelvemib='177
\font\twelvesy=cmsy10 scaled \magstep1     \skewchar\twelvesy='60
\font\twelvesyb=cmbsy10 scaled \magstep1   \skewchar\twelvesyb='60

\font\fourteenrm=cmr10 scaled \magstep2
\font\fourteenbf=cmbx10 scaled \magstep2
\font\fourteenit=cmti10 scaled \magstep2
\font\fourteeni=cmmi10 scaled \magstep2		\skewchar\fourteeni='177
\font\fourteenmib=cmmib10 scaled \magstep2	\skewchar\fourteenmib='177
\font\fourteensy=cmsy10 scaled \magstep2	\skewchar\fourteensy='60
\font\fourteensyb=cmbsy10 scaled \magstep2	\skewchar\fourteensyb='60
\font\fourteensl=cmsl10 scaled \magstep2
\font\fourteentt=cmtt10 scaled \magstep2	\hyphenchar\fourteentt=-1
\font\fourteencsc=cmcsc10 scaled \magstep2
\font\fourteensf=cmss10 scaled \magstep2

\font\seventeenrm=cmr10 scaled \magstep3
\font\seventeenbf=cmbx10 scaled \magstep3
\font\seventeenit=cmti10 scaled \magstep3
\font\seventeeni=cmmi10 scaled \magstep3	\skewchar\seventeeni='177
\font\seventeenmib=cmmib10 scaled \magstep3	\skewchar\seventeenmib='177
\font\seventeensy=cmsy10 scaled \magstep3	\skewchar\seventeensy='60
\font\seventeensyb=cmbsy10 scaled \magstep3	\skewchar\seventeensyb='60
\font\seventeensl=cmsl10 scaled \magstep3
\font\seventeentt=cmtt10 scaled \magstep3	\hyphenchar\seventeentt=-1
\font\seventeencsc=cmcsc10 scaled \magstep3
\font\seventeensf=cmss10 scaled \magstep3
\fi

\def\hexnumber#1{\ifcase#1 0\or1\or2\or3\or4\or5\or6\or7\or8\or9\or
  A\or B\or C\or D\or E\or F\fi}

\ifprod@font
  \edef\@xm{\hexnumber\xmfam}
  \edef\@ym{\hexnumber\ymfam}
\fi

\def\mib{\hexnumber\mibfam}
\def\syb{\hexnumber\sybfam}

\def\makestrut{%
  \setbox\strutbox=\hbox{%
    \vrule height.7\baselineskip depth.3\baselineskip width \z@}%
}

\def\baselinestretch{1}
\newskip\tmp@bls

\def\b@ls#1{
  \tmp@bls=#1\relax
  \baselineskip=#1\relax\makestrut
  \normalbaselineskip=\baselinestretch\tmp@bls
  \normalbaselines
}

\def\nostb@ls#1{
  \normalbaselineskip=#1\relax
  \normalbaselines
  \makestrut
}

%

\newfam\mibfam 
\newfam\sybfam 
\newfam\scfam  
\newfam\sffam  

\def\mit{\fam\@ne}

\def\cal{\fam\tw@}

\def\em{\ifdim\fontdimen1\font>\z@ \rm\else\it\fi}

\textfont3=\tenex
\scriptfont3=\tenex
\scriptscriptfont3=\tenex

\setbox0=\hbox{\tenex B} \p@renwd=\wd0 

\def\eightpoint{
  \def\rm{\fam0\eightrm}%
  \textfont0=\eightrm \scriptfont0=\sixrm \scriptscriptfont0=\fiverm%
  \textfont1=\eighti  \scriptfont1=\sixi  \scriptscriptfont1=\fivei%
  \textfont2=\eightsy \scriptfont2=\sixsy \scriptscriptfont2=\fivesy%
  \textfont\itfam=\eightit\def\it{\fam\itfam\eightit}%
  \ifprod@font
    \scriptfont\itfam=\sixit
      \scriptscriptfont\itfam=\fiveit
  \else
    \scriptfont\itfam=\eightit
      \scriptscriptfont\itfam=\eightit
  \fi
  \textfont\bffam=\eightbf%
    \scriptfont\bffam=\sixbf%
      \scriptscriptfont\bffam=\fivebf%
  \def\bf{\fam\bffam\eightbf}%
  \textfont\slfam=\eightsl\def\sl{\fam\slfam\eightsl}%
  \ifprod@font
    \scriptfont\slfam=\sixsl
      \scriptscriptfont\slfam=\fivesl
  \else
    \scriptfont\slfam=\eightsl
      \scriptscriptfont\slfam=\eightsl
  \fi
  \textfont\ttfam=\eighttt\def\tt{\fam\ttfam\eighttt}%
  \ifprod@font
    \scriptfont\ttfam=\sixtt
      \scriptscriptfont\ttfam=\fivett
  \else
    \scriptfont\ttfam=\eighttt
      \scriptscriptfont\ttfam=\eighttt
  \fi
  \textfont\scfam=\eightcsc\def\sc{\fam\scfam\eightcsc}%
  \ifprod@font
    \scriptfont\scfam=\sixcsc
      \scriptscriptfont\scfam=\fivecsc
  \else
    \scriptfont\scfam=\eightcsc
      \scriptscriptfont\scfam=\eightcsc
  \fi
  \textfont\sffam=\eightsf\def\sf{\fam\sffam\eightsf}%
  \ifprod@font
    \scriptfont\sffam=\sixsf
      \scriptscriptfont\sffam=\fivesf
  \else
    \scriptfont\sffam=\eightsf
      \scriptscriptfont\sffam=\eightsf
  \fi
  \textfont\mibfam=\eightmib
    \scriptfont\mibfam=\sixmib
      \scriptscriptfont\mibfam=\fivemib
  \textfont\sybfam=\eightsyb
    \scriptfont\sybfam=\sixsyb
      \scriptscriptfont\sybfam=\fivesyb
  \ifprod@font
    \textfont\xmfam=\eightxm
      \scriptfont\xmfam=\sixxm
        \scriptscriptfont\xmfam=\fivexm
    \textfont\ymfam=\eightym
      \scriptfont\ymfam=\sixym
        \scriptscriptfont\ymfam=\fiveym
  \fi
  \def\oldstyle{\fam\@ne\eighti}%
  \def\boldstyle{\fam\mibfam\eightmib}%
  \b@ls{10pt}\rm%
}

\def\ninepoint{
  \def\rm{\fam0\ninerm}%
  \textfont0=\ninerm \scriptfont0=\sixrm \scriptscriptfont0=\fiverm%
  \textfont1=\ninei  \scriptfont1=\sixi  \scriptscriptfont1=\fivei%
  \textfont2=\ninesy \scriptfont2=\sixsy \scriptscriptfont2=\fivesy%
  \textfont\itfam=\nineit\def\it{\fam\itfam\nineit}%
  \ifprod@font
    \scriptfont\itfam=\sixit
      \scriptscriptfont\itfam=\fiveit
  \else
    \scriptfont\itfam=\nineit
      \scriptscriptfont\itfam=\nineit
  \fi
  \textfont\bffam=\ninebf%
    \scriptfont\bffam=\sixbf%
      \scriptscriptfont\bffam=\fivebf%
  \def\bf{\fam\bffam\ninebf}%
  \textfont\slfam=\ninesl\def\sl{\fam\slfam\ninesl}%
  \ifprod@font
    \scriptfont\slfam=\sixsl
      \scriptscriptfont\slfam=\fivesl
  \else
    \scriptfont\slfam=\ninesl
      \scriptscriptfont\slfam=\ninesl
  \fi
  \textfont\ttfam=\ninett\def\tt{\fam\ttfam\ninett}%
  \ifprod@font
    \scriptfont\ttfam=\sixtt
      \scriptscriptfont\ttfam=\fivett
  \else
    \scriptfont\ttfam=\ninett
      \scriptscriptfont\ttfam=\ninett
  \fi
  \textfont\scfam=\ninecsc\def\sc{\fam\scfam\ninecsc}%
  \ifprod@font
    \scriptfont\scfam=\sixcsc
      \scriptscriptfont\scfam=\fivecsc
  \else
    \scriptfont\scfam=\ninecsc
      \scriptscriptfont\scfam=\ninecsc
  \fi
  \textfont\sffam=\ninesf\def\sf{\fam\sffam\ninesf}%
  \ifprod@font
    \scriptfont\sffam=\sixsf
      \scriptscriptfont\sffam=\fivesf
  \else
    \scriptfont\sffam=\ninesf
      \scriptscriptfont\sffam=\ninesf
  \fi
  \textfont\mibfam=\ninemib
    \scriptfont\mibfam=\sixmib
      \scriptscriptfont\mibfam=\fivemib
  \textfont\sybfam=\ninesyb
    \scriptfont\sybfam=\sixsyb
      \scriptscriptfont\sybfam=\fivesyb
  \ifprod@font
    \textfont\xmfam=\ninexm
      \scriptfont\xmfam=\sixxm
        \scriptscriptfont\xmfam=\fivexm
    \textfont\ymfam=\nineym
      \scriptfont\ymfam=\sixym
        \scriptscriptfont\ymfam=\fiveym
  \fi
  \def\oldstyle{\fam\@ne\ninei}%
  \def\boldstyle{\fam\mibfam\ninemib}%
  \b@ls{\TextLeading plus \Feathering}\rm%
}

\def\tenpoint{
  \def\rm{\fam0\tenrm}%
  \textfont0=\tenrm \scriptfont0=\sevenrm \scriptscriptfont0=\fiverm%
  \textfont1=\teni  \scriptfont1=\seveni  \scriptscriptfont1=\fivei%
  \textfont2=\tensy \scriptfont2=\sevensy \scriptscriptfont2=\fivesy%
  \textfont\itfam=\tenit\def\it{\fam\itfam\tenit}%
  \ifprod@font
    \scriptfont\itfam=\sevenit
      \scriptscriptfont\itfam=\fiveit
  \else
    \scriptfont\itfam=\tenit
      \scriptscriptfont\itfam=\tenit
  \fi
  \textfont\bffam=\tenbf%
    \scriptfont\bffam=\sevenbf%
      \scriptscriptfont\bffam=\fivebf%
  \def\bf{\fam\bffam\tenbf}%
  \textfont\slfam=\tensl\def\sl{\fam\slfam\tensl}%
  \ifprod@font
    \scriptfont\slfam=\sevensl
      \scriptscriptfont\slfam=\fivesl
  \else
    \scriptfont\slfam=\tensl
      \scriptscriptfont\slfam=\tensl
  \fi
  \textfont\ttfam=\tentt\def\tt{\fam\ttfam\tentt}%
  \ifprod@font
    \scriptfont\ttfam=\seventt
      \scriptscriptfont\ttfam=\fivett
  \else
    \scriptfont\ttfam=\tentt
      \scriptscriptfont\ttfam=\tentt
  \fi
  \textfont\scfam=\tencsc\def\sc{\fam\scfam\tencsc}%
  \ifprod@font
    \scriptfont\scfam=\sevencsc
      \scriptscriptfont\scfam=\fivecsc
  \else
    \scriptfont\scfam=\tencsc
      \scriptscriptfont\scfam=\tencsc
  \fi
  \textfont\sffam=\tensf\def\sf{\fam\sffam\tensf}%
  \ifprod@font
    \scriptfont\sffam=\sevensf
      \scriptscriptfont\sffam=\fivesf
  \else
    \scriptfont\sffam=\tensf
      \scriptscriptfont\sffam=\tensf
  \fi
  \textfont\mibfam=\tenmib
    \scriptfont\mibfam=\sevenmib
      \scriptscriptfont\mibfam=\fivemib
  \textfont\sybfam=\tensyb
    \scriptfont\sybfam=\sevensyb
      \scriptscriptfont\sybfam=\fivesyb
  \ifprod@font
    \textfont\xmfam=\tenxm
      \scriptfont\xmfam=\sevenxm
        \scriptscriptfont\xmfam=\fivexm
    \textfont\ymfam=\tenym
      \scriptfont\ymfam=\sevenym
        \scriptscriptfont\ymfam=\fiveym
  \fi
  \def\oldstyle{\fam\@ne\teni}%
  \def\boldstyle{\fam\mibfam\tenmib}%
  \b@ls{11pt}\rm%
}

\def\elevenpoint{
  \def\rm{\fam0\elevenrm}%
  \textfont0=\elevenrm \scriptfont0=\eightrm \scriptscriptfont0=\sixrm%
  \textfont1=\eleveni  \scriptfont1=\eighti  \scriptscriptfont1=\sixi%
  \textfont2=\elevensy \scriptfont2=\eightsy \scriptscriptfont2=\sixsy%
  \textfont\itfam=\elevenit\def\it{\fam\itfam\elevenit}%
  \ifprod@font
    \scriptfont\itfam=\eightit
      \scriptscriptfont\itfam=\sixit
  \else
    \scriptfont\itfam=\elevenit
      \scriptscriptfont\itfam=\elevenit
  \fi
  \textfont\bffam=\elevenbf%
    \scriptfont\bffam=\eightbf%
      \scriptscriptfont\bffam=\sixbf%
  \def\bf{\fam\bffam\elevenbf}%
  \textfont\slfam=\elevensl\def\sl{\fam\slfam\elevensl}%
  \ifprod@font
    \scriptfont\slfam=\eightsl
      \scriptscriptfont\slfam=\sixsl
  \else
    \scriptfont\slfam=\elevensl
      \scriptscriptfont\slfam=\elevensl
  \fi
  \textfont\ttfam=\eleventt\def\tt{\fam\ttfam\eleventt}%
  \ifprod@font
    \scriptfont\ttfam=\eighttt
      \scriptscriptfont\ttfam=\sixtt
  \else
    \scriptfont\ttfam=\eleventt
      \scriptscriptfont\ttfam=\eleventt
  \fi
  \textfont\scfam=\elevencsc\def\sc{\fam\scfam\elevencsc}%
  \ifprod@font
    \scriptfont\scfam=\eightcsc
      \scriptscriptfont\scfam=\sixcsc
  \else
    \scriptfont\scfam=\elevencsc
      \scriptscriptfont\scfam=\elevencsc
  \fi
  \textfont\sffam=\elevensf\def\sf{\fam\sffam\elevensf}%
  \ifprod@font
    \scriptfont\sffam=\eightsf
      \scriptscriptfont\sffam=\sixsf
  \else
    \scriptfont\sffam=\elevensf
      \scriptscriptfont\sffam=\elevensf
  \fi
  \textfont\mibfam=\elevenmib
    \scriptfont\mibfam=\eightmib
      \scriptscriptfont\mibfam=\sixmib
  \textfont\sybfam=\elevensyb
    \scriptfont\sybfam=\eightsyb
      \scriptscriptfont\sybfam=\sixsyb
  \ifprod@font
    \textfont\xmfam=\elevenxm
      \scriptfont\xmfam=\eightxm
       \scriptscriptfont\xmfam=\sixxm
    \textfont\ymfam=\elevenym
      \scriptfont\ymfam=\eightym
        \scriptscriptfont\ymfam=\sixym
   \fi
  \def\oldstyle{\fam\@ne\eleveni}%
  \def\boldstyle{\fam\mibfam\elevenmib}%
  \b@ls{13pt}\rm%
}

\def\fourteenpoint{
  \def\rm{\fam0\fourteenrm}%
  \textfont0\fourteenrm  \scriptfont0\tenrm  \scriptscriptfont0\sevenrm%
  \textfont1\fourteeni   \scriptfont1\teni   \scriptscriptfont1\seveni%
  \textfont2\fourteensy  \scriptfont2\tensy  \scriptscriptfont2\sevensy%
  \textfont\itfam=\fourteenit\def\it{\fam\itfam\fourteenit}%
  \ifprod@font
    \scriptfont\itfam=\tenit
      \scriptscriptfont\itfam=\sevenit
  \else
    \scriptfont\itfam=\fourteenit
      \scriptscriptfont\itfam=\fourteenit
  \fi
  \textfont\bffam=\fourteenbf%
    \scriptfont\bffam=\tenbf%
      \scriptscriptfont\bffam=\sevenbf%
  \def\bf{\fam\bffam\fourteenbf}%
  \textfont\slfam=\fourteensl\def\sl{\fam\slfam\fourteensl}%
  \ifprod@font
    \scriptfont\slfam=\tensl
      \scriptscriptfont\slfam=\sevensl
  \else
    \scriptfont\slfam=\fourteensl
      \scriptscriptfont\slfam=\fourteensl
  \fi
  \textfont\ttfam=\fourteentt\def\tt{\fam\ttfam\fourteentt}%
  \ifprod@font
    \scriptfont\ttfam=\tentt
      \scriptscriptfont\ttfam=\seventt
  \else
    \scriptfont\ttfam=\fourteentt
      \scriptscriptfont\ttfam=\fourteentt
  \fi
  \textfont\scfam=\fourteencsc\def\sc{\fam\scfam\fourteencsc}%
  \ifprod@font
    \scriptfont\scfam=\tencsc
      \scriptscriptfont\scfam=\sevencsc
  \else
    \scriptfont\scfam=\fourteencsc
      \scriptscriptfont\scfam=\fourteencsc
  \fi
  \textfont\sffam=\fourteensf\def\sf{\fam\sffam\fourteensf}%
  \ifprod@font
    \scriptfont\sffam=\tensf
      \scriptscriptfont\sffam=\sevensf
  \else
    \scriptfont\sffam=\fourteensf
      \scriptscriptfont\sffam=\fourteensf
  \fi
  \textfont\mibfam=\fourteenmib
    \scriptfont\mibfam=\tenmib
      \scriptscriptfont\mibfam=\sevenmib
  \textfont\sybfam=\fourteensyb
    \scriptfont\sybfam=\tensyb
      \scriptscriptfont\sybfam=\sevensyb
  \ifprod@font
    \textfont\xmfam=\fourteenxm
      \scriptfont\xmfam=\tenxm
        \scriptscriptfont\xmfam=\sevenxm
   \textfont\ymfam=\fourteenym
      \scriptfont\ymfam=\tenym
        \scriptscriptfont\ymfam=\sevenym
  \fi
  \def\oldstyle{\fam\@ne\fourteeni}%
  \def\boldstyle{\fam\mibfam\fourteenmib}%
  \b@ls{17pt}\rm%
}

\def\seventeenpoint{
  \def\rm{\fam0\seventeenrm}%
  \textfont0\seventeenrm  \scriptfont0\twelverm  \scriptscriptfont0\tenrm%
  \textfont1\seventeeni   \scriptfont1\twelvei   \scriptscriptfont1\teni%
  \textfont2\seventeensy  \scriptfont2\twelvesy  \scriptscriptfont2\tensy%
  \textfont\itfam=\seventeenit\def\it{\fam\itfam\seventeenit}%
  \ifprod@font
    \scriptfont\itfam=\twelveit
      \scriptscriptfont\itfam=\tenit
  \else
    \scriptfont\itfam=\seventeenit
      \scriptscriptfont\itfam=\seventeenit
  \fi
  \textfont\bffam=\seventeenbf%
    \scriptfont\bffam=\twelvebf%
      \scriptscriptfont\bffam=\tenbf%
  \def\bf{\fam\bffam\seventeenbf}%
  \textfont\slfam=\seventeensl\def\sl{\fam\slfam\seventeensl}%
  \ifprod@font
    \scriptfont\slfam=\twelvesl
      \scriptscriptfont\slfam=\tensl
  \else
    \scriptfont\slfam=\seventeensl
      \scriptscriptfont\slfam=\seventeensl
  \fi
  \textfont\ttfam=\seventeentt\def\tt{\fam\ttfam\seventeentt}%
  \ifprod@font
    \scriptfont\ttfam=\twelvett
      \scriptscriptfont\ttfam=\tentt
  \else
    \scriptfont\ttfam=\seventeentt
      \scriptscriptfont\ttfam=\seventeentt
  \fi
  \textfont\scfam=\seventeencsc\def\sc{\fam\scfam\seventeencsc}%
  \ifprod@font
    \scriptfont\scfam=\twelvecsc
      \scriptscriptfont\scfam=\tencsc
  \else
    \scriptfont\scfam=\seventeencsc
      \scriptscriptfont\scfam=\seventeencsc
  \fi
  \textfont\sffam=\seventeensf\def\sf{\fam\sffam\seventeensf}%
  \ifprod@font
    \scriptfont\sffam=\twelvesf
      \scriptscriptfont\sffam=\tensf
  \else
    \scriptfont\sffam=\seventeensf
      \scriptscriptfont\sffam=\seventeensf
  \fi
  \textfont\mibfam=\seventeenmib
    \scriptfont\mibfam=\twelvemib
      \scriptscriptfont\mibfam=\tenmib
  \textfont\sybfam=\seventeensyb
    \scriptfont\sybfam=\twelvesyb
      \scriptscriptfont\sybfam=\tensyb
  \ifprod@font
    \textfont\xmfam=\seventeenxm
      \scriptfont\xmfam=\twelvexm
        \scriptscriptfont\xmfam=\tenxm
    \textfont\ymfam=\seventeenym
      \scriptfont\ymfam=\twelveym
        \scriptscriptfont\ymfam=\tenym
  \fi
  \def\oldstyle{\fam\@ne\seventeeni}%
  \def\boldstyle{\fam\mibfam\seventeenmib}%
  \b@ls{20pt}\rm%
}

\lineskip=1pt      \normallineskip=\lineskip
\lineskiplimit=\z@ \normallineskiplimit=\lineskiplimit


\def\loadboldmathnames{%
  \mathchardef\balpha="0\mib0B
  \mathchardef\bbeta="0\mib0C
  \mathchardef\bgamma="0\mib0D
  \mathchardef\bdelta="0\mib0E
  \mathchardef\bepsilon="0\mib0F
  \mathchardef\bzeta="0\mib10
  \mathchardef\boldeta="0\mib11 
  \mathchardef\btheta="0\mib12
  \mathchardef\biota="0\mib13
  \mathchardef\bkappa="0\mib14
  \mathchardef\blambda="0\mib15
  \mathchardef\bmu="0\mib16
  \mathchardef\bnu="0\mib17
  \mathchardef\bxi="0\mib18
  \mathchardef\bpi="0\mib19
  \mathchardef\brho="0\mib1A
  \mathchardef\bsigma="0\mib1B
  \mathchardef\btau="0\mib1C
  \mathchardef\bupsilon="0\mib1D
  \mathchardef\bphi="0\mib1E
  \mathchardef\bchi="0\mib1F
  \mathchardef\bpsi="0\mib20
  \mathchardef\bomega="0\mib21
  \mathchardef\bvarepsilon="0\mib22
  \mathchardef\bvartheta="0\mib23
  \mathchardef\bvarpi="0\mib24
  \mathchardef\bvarrho="0\mib25
  \mathchardef\bvarsigma="0\mib26
  \mathchardef\bvarphi="0\mib27
  \mathchardef\baleph="0\syb40
  \mathchardef\bimath="0\mib7B
  \mathchardef\bjmath="0\mib7C
  \mathchardef\bell="0\mib60
  \mathchardef\bwp="0\mib7D
  \mathchardef\bRe="0\syb3C
  \mathchardef\bIm="0\syb3D
  \mathchardef\bpartial="0\mib40
  \mathchardef\binfty="0\syb31
  \mathchardef\bprime="0\syb30
  \mathchardef\bemptyset="0\syb3B
  \mathchardef\bnabla="0\syb72
  \mathchardef\btop="0\syb3E
  \mathchardef\bbot="0\syb3F
  \mathchardef\btriangle="0\syb34
  \mathchardef\bforall="0\syb38
  \mathchardef\bexists="0\syb39
  \mathchardef\bneg="0\syb3A
  \mathchardef\bflat="0\mib5B
  \mathchardef\bnatural="0\mib5C
  \mathchardef\bsharp="0\mib5D
  \mathchardef\bclubsuit="0\syb7C
  \mathchardef\bdiamondsuit="0\syb7D
  \mathchardef\bheartsuit="0\syb7E
  \mathchardef\bspadesuit="0\syb7F
  \mathchardef\bsmallint="1\syb73
  \mathchardef\btriangleleft="2\mib2F
  \mathchardef\btriangleright="2\mib2E
  \mathchardef\bbigtriangleup="2\syb34
  \mathchardef\bbigtriangledown="2\syb35
  \mathchardef\bwedge="2\syb5E
  \mathchardef\bvee="2\syb5F
  \mathchardef\bcap="2\syb5C
  \mathchardef\bcup="2\syb5B
  \mathchardef\bddagger="2\syb7A
  \mathchardef\bdagger="2\syb79
  \mathchardef\bsqcap="2\syb75
  \mathchardef\bsqcup="2\syb74
  \mathchardef\buplus="2\syb5D
  \mathchardef\bamalg="2\syb71
  \mathchardef\bdiamond="2\syb05
  \mathchardef\bbullet="2\syb0F
  \mathchardef\bwr="2\syb6F
  \mathchardef\bdiv="2\syb04
  \mathchardef\bodot="2\syb0C
  \mathchardef\boslash="2\syb0B
  \mathchardef\botimes="2\syb0A
  \mathchardef\bominus="2\syb09
  \mathchardef\boplus="2\syb08
  \mathchardef\bmp="2\syb07
  \mathchardef\bpm="2\syb06
  \mathchardef\bcirc="2\syb0E
  \mathchardef\bbigcirc="2\syb0D
  \mathchardef\bsetminus="2\syb6E
  \mathchardef\bcdot="2\syb01
  \mathchardef\bast="2\syb03
  \mathchardef\btimes="2\syb02
  \mathchardef\bstar="2\mib3F
  \mathchardef\bpropto="3\syb2F
  \mathchardef\bsqsubseteq="3\syb76
  \mathchardef\bsqsupseteq="3\syb77
  \mathchardef\bparallel="3\syb6B
  \mathchardef\bmid="3\syb6A
  \mathchardef\bdashv="3\syb61
  \mathchardef\bvdash="3\syb60
  \mathchardef\bnearrow="3\syb25
  \mathchardef\bsearrow="3\syb26
  \mathchardef\bnwarrow="3\syb2D
  \mathchardef\bswarrow="3\syb2E
  \mathchardef\bLeftrightarrow="3\syb2C
  \mathchardef\bLeftarrow="3\syb28
  \mathchardef\bRightarrow="3\syb29
  \mathchardef\bleq="3\syb14
  \mathchardef\bgeq="3\syb15
  \mathchardef\bsucc="3\syb1F
  \mathchardef\bprec="3\syb1E
  \mathchardef\bapprox="3\syb19
  \mathchardef\bsucceq="3\syb17
  \mathchardef\bpreceq="3\syb16
  \mathchardef\bsupset="3\syb1B
  \mathchardef\bsubset="3\syb1A
  \mathchardef\bsupseteq="3\syb13
  \mathchardef\bsubseteq="3\syb12
  \mathchardef\bin="3\syb32
  \mathchardef\bni="3\syb33
  \mathchardef\bgg="3\syb1D
  \mathchardef\bll="3\syb1C
  \mathchardef\bnot="3\syb36
  \mathchardef\bleftrightarrow="3\syb24
  \mathchardef\bleftarrow="3\syb20
  \mathchardef\brightarrow="3\syb21
  \mathchardef\bmapstochar="3\syb37
  \mathchardef\bsim="3\syb18
  \mathchardef\bsimeq="3\syb27
  \mathchardef\bperp="3\syb3F
  \mathchardef\bequiv="3\syb11
  \mathchardef\basymp="3\syb10
  \mathchardef\bsmile="3\mib5E
  \mathchardef\bfrown="3\mib5F
  \mathchardef\bleftharpoonup="3\mib28
  \mathchardef\bleftharpoondown="3\mib29
  \mathchardef\brightharpoonup="3\mib2A
  \mathchardef\brightharpoondown="3\mib2B
  \mathchardef\blhook="3\mib2C
  \mathchardef\brhook="3\mib2D
  \mathchardef\bldotp="6\mib3A
  \mathchardef\bcdotp="6\syb01
}


\def\la{\mathrel{\mathchoice {\vcenter{\offinterlineskip\halign{\hfil
$\displaystyle##$\hfil\cr<\cr\sim\cr}}}
{\vcenter{\offinterlineskip\halign{\hfil$\textstyle##$\hfil\cr
<\cr\sim\cr}}}
{\vcenter{\offinterlineskip\halign{\hfil$\scriptstyle##$\hfil\cr
<\cr\sim\cr}}}
{\vcenter{\offinterlineskip\halign{\hfil$\scriptscriptstyle##$\hfil\cr
<\cr\sim\cr}}}}}

\def\ga{\mathrel{\mathchoice {\vcenter{\offinterlineskip\halign{\hfil
$\displaystyle##$\hfil\cr>\cr\sim\cr}}}
{\vcenter{\offinterlineskip\halign{\hfil$\textstyle##$\hfil\cr
>\cr\sim\cr}}}
{\vcenter{\offinterlineskip\halign{\hfil$\scriptstyle##$\hfil\cr
>\cr\sim\cr}}}
{\vcenter{\offinterlineskip\halign{\hfil$\scriptscriptstyle##$\hfil\cr
>\cr\sim\cr}}}}}

\def\getsto{\mathrel{\mathchoice {\vcenter{\offinterlineskip
\halign{\hfil
$\displaystyle##$\hfil\cr\gets\cr\to\cr}}}
{\vcenter{\offinterlineskip\halign{\hfil$\textstyle##$\hfil\cr\gets
\cr\to\cr}}}
{\vcenter{\offinterlineskip\halign{\hfil$\scriptstyle##$\hfil\cr\gets
\cr\to\cr}}}
{\vcenter{\offinterlineskip\halign{\hfil$\scriptscriptstyle##$\hfil\cr
\gets\cr\to\cr}}}}}

\def\lid{\mathrel{\mathchoice {\vcenter{\offinterlineskip\halign{\hfil
$\displaystyle##$\hfil\cr<\cr\noalign{\vskip1.2pt}=\cr}}}
{\vcenter{\offinterlineskip\halign{\hfil$\textstyle##$\hfil\cr<\cr
\noalign{\vskip1.2pt}=\cr}}}
{\vcenter{\offinterlineskip\halign{\hfil$\scriptstyle##$\hfil\cr<\cr
\noalign{\vskip1pt}=\cr}}}
{\vcenter{\offinterlineskip\halign{\hfil$\scriptscriptstyle##$\hfil\cr
<\cr
\noalign{\vskip0.9pt}=\cr}}}}}

\def\gid{\mathrel{\mathchoice {\vcenter{\offinterlineskip\halign{\hfil
$\displaystyle##$\hfil\cr>\cr\noalign{\vskip1.2pt}=\cr}}}
{\vcenter{\offinterlineskip\halign{\hfil$\textstyle##$\hfil\cr>\cr
\noalign{\vskip1.2pt}=\cr}}}
{\vcenter{\offinterlineskip\halign{\hfil$\scriptstyle##$\hfil\cr>\cr
\noalign{\vskip1pt}=\cr}}}
{\vcenter{\offinterlineskip\halign{\hfil$\scriptscriptstyle##$\hfil\cr
>\cr
\noalign{\vskip0.9pt}=\cr}}}}}

\def\grole{\mathrel{\mathchoice {\vcenter{\offinterlineskip\halign{\hfil
$\displaystyle##$\hfil\cr>\cr\noalign{\vskip-1.5pt}<\cr}}}
{\vcenter{\offinterlineskip\halign{\hfil$\textstyle##$\hfil\cr
>\cr\noalign{\vskip-1.5pt}<\cr}}}
{\vcenter{\offinterlineskip\halign{\hfil$\scriptstyle##$\hfil\cr
>\cr\noalign{\vskip-1pt}<\cr}}}
{\vcenter{\offinterlineskip\halign{\hfil$\scriptscriptstyle##$\hfil\cr
>\cr\noalign{\vskip-0.5pt}<\cr}}}}}

\def\leogr{\mathrel{\mathchoice {\vcenter{\offinterlineskip\halign{\hfil
$\displaystyle##$\hfil\cr<\cr\noalign{\vskip-1.5pt}>\cr}}}
{\vcenter{\offinterlineskip\halign{\hfil$\textstyle##$\hfil\cr
<\cr\noalign{\vskip-1.5pt}>\cr}}}
{\vcenter{\offinterlineskip\halign{\hfil$\scriptstyle##$\hfil\cr
<\cr\noalign{\vskip-1pt}>\cr}}}
{\vcenter{\offinterlineskip\halign{\hfil$\scriptscriptstyle##$\hfil\cr
<\cr\noalign{\vskip-0.5pt}>\cr}}}}}

\def\loa{\mathrel{\mathchoice {\vcenter{\offinterlineskip\halign{\hfil
$\displaystyle##$\hfil\cr<\cr\approx\cr}}}
{\vcenter{\offinterlineskip\halign{\hfil$\textstyle##$\hfil\cr
<\cr\approx\cr}}}
{\vcenter{\offinterlineskip\halign{\hfil$\scriptstyle##$\hfil\cr
<\cr\approx\cr}}}
{\vcenter{\offinterlineskip\halign{\hfil$\scriptscriptstyle##$\hfil\cr
<\cr\approx\cr}}}}}

\def\goa{\mathrel{\mathchoice {\vcenter{\offinterlineskip\halign{\hfil
$\displaystyle##$\hfil\cr>\cr\approx\cr}}}
{\vcenter{\offinterlineskip\halign{\hfil$\textstyle##$\hfil\cr
>\cr\approx\cr}}}
{\vcenter{\offinterlineskip\halign{\hfil$\scriptstyle##$\hfil\cr
>\cr\approx\cr}}}
{\vcenter{\offinterlineskip\halign{\hfil$\scriptscriptstyle##$\hfil\cr
>\cr\approx\cr}}}}}

\def\diameter{{\ifmmode\mathchoice
{\ooalign{\hfil\hbox{$\displaystyle/$}\hfil\crcr
{\hbox{$\displaystyle\mathchar"20D$}}}}
{\ooalign{\hfil\hbox{$\textstyle/$}\hfil\crcr
{\hbox{$\textstyle\mathchar"20D$}}}}
{\ooalign{\hfil\hbox{$\scriptstyle/$}\hfil\crcr
{\hbox{$\scriptstyle\mathchar"20D$}}}}
{\ooalign{\hfil\hbox{$\scriptscriptstyle/$}\hfil\crcr
{\hbox{$\scriptscriptstyle\mathchar"20D$}}}}
\else{\ooalign{\hfil/\hfil\crcr\mathhexbox20D}}%
\fi}}

\def\sq{\ifmmode\squareforqed\else{\unskip\nobreak\hfil
\penalty50\hskip1em\null\nobreak\hfil\squareforqed
\parfillskip=0pt\finalhyphendemerits=0\endgraf}\fi}
\def\squareforqed{\hbox{\rlap{$\sqcap$}$\sqcup$}}


\def\bbbc{{\mathchoice {\setbox0=\hbox{$\displaystyle\rm C$}\hbox{\hbox
to0pt{\kern0.4\wd0\vrule height0.9\ht0\hss}\box0}}
{\setbox0=\hbox{$\textstyle\rm C$}\hbox{\hbox
to0pt{\kern0.4\wd0\vrule height0.9\ht0\hss}\box0}}
{\setbox0=\hbox{$\scriptstyle\rm C$}\hbox{\hbox
to0pt{\kern0.4\wd0\vrule height0.9\ht0\hss}\box0}}
{\setbox0=\hbox{$\scriptscriptstyle\rm C$}\hbox{\hbox
to0pt{\kern0.4\wd0\vrule height0.9\ht0\hss}\box0}}}}
\def\bbbq{{\mathchoice {\setbox0=\hbox{$\displaystyle\rm
Q$}\hbox{\raise
0.15\ht0\hbox to0pt{\kern0.4\wd0\vrule height0.8\ht0\hss}\box0}}
{\setbox0=\hbox{$\textstyle\rm Q$}\hbox{\raise
0.15\ht0\hbox to0pt{\kern0.4\wd0\vrule height0.8\ht0\hss}\box0}}
{\setbox0=\hbox{$\scriptstyle\rm Q$}\hbox{\raise
0.15\ht0\hbox to0pt{\kern0.4\wd0\vrule height0.7\ht0\hss}\box0}}
{\setbox0=\hbox{$\scriptscriptstyle\rm Q$}\hbox{\raise
0.15\ht0\hbox to0pt{\kern0.4\wd0\vrule height0.7\ht0\hss}\box0}}}}
\def\bbbt{{\mathchoice {\setbox0=\hbox{$\displaystyle\rm
T$}\hbox{\hbox to0pt{\kern0.3\wd0\vrule height0.9\ht0\hss}\box0}}
{\setbox0=\hbox{$\textstyle\rm T$}\hbox{\hbox
to0pt{\kern0.3\wd0\vrule height0.9\ht0\hss}\box0}}
{\setbox0=\hbox{$\scriptstyle\rm T$}\hbox{\hbox
to0pt{\kern0.3\wd0\vrule height0.9\ht0\hss}\box0}}
{\setbox0=\hbox{$\scriptscriptstyle\rm T$}\hbox{\hbox
to0pt{\kern0.3\wd0\vrule height0.9\ht0\hss}\box0}}}}
\def\bbbs{{\mathchoice
{\setbox0=\hbox{$\displaystyle     \rm S$}\hbox{\raise0.5\ht0\hbox
to0pt{\kern0.35\wd0\vrule height0.45\ht0\hss}\hbox
to0pt{\kern0.55\wd0\vrule height0.5\ht0\hss}\box0}}
{\setbox0=\hbox{$\textstyle        \rm S$}\hbox{\raise0.5\ht0\hbox
to0pt{\kern0.35\wd0\vrule height0.45\ht0\hss}\hbox
to0pt{\kern0.55\wd0\vrule height0.5\ht0\hss}\box0}}
{\setbox0=\hbox{$\scriptstyle      \rm S$}\hbox{\raise0.5\ht0\hbox
to0pt{\kern0.35\wd0\vrule height0.45\ht0\hss}\raise0.05\ht0\hbox
to0pt{\kern0.5\wd0\vrule height0.45\ht0\hss}\box0}}
{\setbox0=\hbox{$\scriptscriptstyle\rm S$}\hbox{\raise0.5\ht0\hbox
to0pt{\kern0.4\wd0\vrule height0.45\ht0\hss}\raise0.05\ht0\hbox
to0pt{\kern0.55\wd0\vrule height0.45\ht0\hss}\box0}}}}
\def\bbbz{{\mathchoice {\hbox{$\sf\textstyle Z\kern-0.4em Z$}}
{\hbox{$\sf\textstyle Z\kern-0.4em Z$}}
{\hbox{$\sf\scriptstyle Z\kern-0.3em Z$}}
{\hbox{$\sf\scriptscriptstyle Z\kern-0.2em Z$}}}}


\ifprod@font
  \mathchardef\la="3\@xm2E
  \mathchardef\getsto="3\@xm1C
  \mathchardef\lid="3\@xm35
  \mathchardef\grole="3\@xm3F
  \mathchardef\loa="3\@xm2F
  \mathchardef\ga="3\@xm26
  \mathchardef\gid="3\@xm3D
  \mathchardef\leogr="3\@xm37
  \mathchardef\goa="3\@xm27
  \mathchardef\sq="0\@xm03
%
%
\def\diameter{{%
  \ifmmode
    \mathchoice
    {\ooalign{\hfil\hbox{$\displaystyle/$}\hfil\crcr
    {\lower.2ex\hbox{$\displaystyle\mathchar"20D$}}}}%
    {\ooalign{\hfil\hbox{$\textstyle/$}\hfil\crcr
    {\lower.2ex\hbox{$\textstyle\mathchar"20D$}}}}%
    {\ooalign{\hfil\hbox{$\scriptstyle/$}\hfil\crcr
    {\lower.1ex\hbox{$\scriptstyle\mathchar"20D$}}}}%
    {\ooalign{\hfil\hbox{$\scriptscriptstyle/$}\hfil\crcr
    {\lower.1ex\hbox{$\scriptscriptstyle\mathchar"20D$}}}}%
  \else
    {\ooalign{\hfil/\hfil\crcr\lower.2ex\hbox{\mathhexbox20D}}}%
  \fi
}}
%
%

\def\bbbc{{\Bbb{C}}}
\def\bbbq{{\Bbb{Q}}}
\def\bbbt{{\Bbb{T}}}
\def\bbbs{{\Bbb{S}}}
\def\bbbz{{\Bbb{Z}}}
\fi


\ifprod@font
\mathchardef\boxdot="2\@xm00
\mathchardef\boxplus="2\@xm01
\mathchardef\boxtimes="2\@xm02
\mathchardef\square="0\@xm03
\mathchardef\blacksquare="0\@xm04
\mathchardef\centerdot="2\@xm05
\mathchardef\lozenge="0\@xm06
\mathchardef\blacklozenge="0\@xm07
\mathchardef\circlearrowright="3\@xm08
\mathchardef\circlearrowleft="3\@xm09
\mathchardef\rightleftharpoons="3\@xm0A
\mathchardef\leftrightharpoons="3\@xm0B
\mathchardef\boxminus="2\@xm0C
\mathchardef\Vdash="3\@xm0D
\mathchardef\Vvdash="3\@xm0E
\mathchardef\vDash="3\@xm0F
\mathchardef\twoheadrightarrow="3\@xm10
\mathchardef\twoheadleftarrow="3\@xm11
\mathchardef\leftleftarrows="3\@xm12
\mathchardef\rightrightarrows="3\@xm13
\mathchardef\upuparrows="3\@xm14
\mathchardef\downdownarrows="3\@xm15
\mathchardef\upharpoonright="3\@xm16

\mathchardef\downharpoonright="3\@xm17
\mathchardef\upharpoonleft="3\@xm18
\mathchardef\downharpoonleft="3\@xm19
\mathchardef\rightarrowtail="3\@xm1A
\mathchardef\leftarrowtail="3\@xm1B
\mathchardef\leftrightarrows="3\@xm1C
\mathchardef\rightleftarrows="3\@xm1D
\mathchardef\Lsh="3\@xm1E
\mathchardef\Rsh="3\@xm1F
\mathchardef\rightsquigarrow="3\@xm20
\mathchardef\leftrightsquigarrow="3\@xm21
\mathchardef\looparrowleft="3\@xm22
\mathchardef\looparrowright="3\@xm23
\mathchardef\circeq="3\@xm24
\mathchardef\succsim="3\@xm25
\mathchardef\gtrsim="3\@xm26
\mathchardef\gtrapprox="3\@xm27
\mathchardef\multimap="3\@xm28
\mathchardef\therefore="3\@xm29
\mathchardef\because="3\@xm2A
\mathchardef\doteqdot="3\@xm2B

\mathchardef\triangleq="3\@xm2C
\mathchardef\precsim="3\@xm2D
\mathchardef\lesssim="3\@xm2E
\mathchardef\lessapprox="3\@xm2F
\mathchardef\eqslantless="3\@xm30
\mathchardef\eqslantgtr="3\@xm31
\mathchardef\curlyeqprec="3\@xm32
\mathchardef\curlyeqsucc="3\@xm33
\mathchardef\preccurlyeq="3\@xm34
\mathchardef\leqq="3\@xm35
\mathchardef\leqslant="3\@xm36
\mathchardef\lessgtr="3\@xm37
\mathchardef\backprime="0\@xm38
\mathchardef\risingdotseq="3\@xm3A
\mathchardef\fallingdotseq="3\@xm3B
\mathchardef\succcurlyeq="3\@xm3C
\mathchardef\geqq="3\@xm3D
\mathchardef\geqslant="3\@xm3E
\mathchardef\gtrless="3\@xm3F
\mathchardef\sqsubset="3\@xm40
\mathchardef\sqsupset="3\@xm41
\mathchardef\vartriangleright="3\@xm42
\mathchardef\vartriangleleft="3\@xm43
\mathchardef\trianglerighteq="3\@xm44
\mathchardef\trianglelefteq="3\@xm45
\mathchardef\bigstar="0\@xm46
\mathchardef\between="3\@xm47
\mathchardef\blacktriangledown="0\@xm48
\mathchardef\blacktriangleright="3\@xm49
\mathchardef\blacktriangleleft="3\@xm4A
\mathchardef\vartriangle="0\@xm4D
\mathchardef\blacktriangle="0\@xm4E
\mathchardef\triangledown="0\@xm4F
\mathchardef\eqcirc="3\@xm50
\mathchardef\lesseqgtr="3\@xm51
\mathchardef\gtreqless="3\@xm52
\mathchardef\lesseqqgtr="3\@xm53
\mathchardef\gtreqqless="3\@xm54
\mathchardef\Rrightarrow="3\@xm56
\mathchardef\Lleftarrow="3\@xm57
\mathchardef\veebar="2\@xm59
\mathchardef\barwedge="2\@xm5A
\mathchardef\doublebarwedge="2\@xm5B
\mathchardef\angle="0\@xm5C
\mathchardef\measuredangle="0\@xm5D
\mathchardef\sphericalangle="0\@xm5E
\mathchardef\varpropto="3\@xm5F
\mathchardef\smallsmile="3\@xm60
\mathchardef\smallfrown="3\@xm61
\mathchardef\Subset="3\@xm62
\mathchardef\Supset="3\@xm63
\mathchardef\Cup="2\@xm64

\mathchardef\Cap="2\@xm65

\mathchardef\curlywedge="2\@xm66
\mathchardef\curlyvee="2\@xm67
\mathchardef\leftthreetimes="2\@xm68
\mathchardef\rightthreetimes="2\@xm69
\mathchardef\subseteqq="3\@xm6A
\mathchardef\supseteqq="3\@xm6B
\mathchardef\bumpeq="3\@xm6C
\mathchardef\Bumpeq="3\@xm6D
\mathchardef\lll="3\@xm6E

\mathchardef\ggg="3\@xm6F

\mathchardef\circledS="0\@xm73
\mathchardef\pitchfork="3\@xm74
\mathchardef\dotplus="2\@xm75
\mathchardef\backsim="3\@xm76
\mathchardef\backsimeq="3\@xm77
\mathchardef\complement="0\@xm7B
\mathchardef\intercal="2\@xm7C
\mathchardef\circledcirc="2\@xm7D
\mathchardef\circledast="2\@xm7E
\mathchardef\circleddash="2\@xm7F
\def\ulcorner{\delimiter"4\@xm70\@xm70 }
\def\urcorner{\delimiter"5\@xm71\@xm71 }
\def\llcorner{\delimiter"4\@xm78\@xm78 }
\def\lrcorner{\delimiter"5\@xm79\@xm79 }
\def\yen{\mathhexbox\@xm55 }
\def\checkmark{\mathhexbox\@xm58 }
\def\circledR{\mathhexbox\@xm72 }
\def\maltese{\mathhexbox\@xm7A }
\mathchardef\lvertneqq="3\@ym00
\mathchardef\gvertneqq="3\@ym01
\mathchardef\nleq="3\@ym02
\mathchardef\ngeq="3\@ym03
\mathchardef\nless="3\@ym04
\mathchardef\ngtr="3\@ym05
\mathchardef\nprec="3\@ym06
\mathchardef\nsucc="3\@ym07
\mathchardef\lneqq="3\@ym08
\mathchardef\gneqq="3\@ym09
\mathchardef\nleqslant="3\@ym0A
\mathchardef\ngeqslant="3\@ym0B
\mathchardef\lneq="3\@ym0C
\mathchardef\gneq="3\@ym0D
\mathchardef\npreceq="3\@ym0E
\mathchardef\nsucceq="3\@ym0F
\mathchardef\precnsim="3\@ym10
\mathchardef\succnsim="3\@ym11
\mathchardef\lnsim="3\@ym12
\mathchardef\gnsim="3\@ym13
\mathchardef\nleqq="3\@ym14
\mathchardef\ngeqq="3\@ym15
\mathchardef\precneqq="3\@ym16
\mathchardef\succneqq="3\@ym17
\mathchardef\precnapprox="3\@ym18
\mathchardef\succnapprox="3\@ym19
\mathchardef\lnapprox="3\@ym1A
\mathchardef\gnapprox="3\@ym1B
\mathchardef\nsim="3\@ym1C
\mathchardef\ncong="3\@ym1D

\mathchardef\varsubsetneq="3\@ym20
\mathchardef\varsupsetneq="3\@ym21
\mathchardef\nsubseteqq="3\@ym22
\mathchardef\nsupseteqq="3\@ym23
\mathchardef\subsetneqq="3\@ym24
\mathchardef\supsetneqq="3\@ym25
\mathchardef\varsubsetneqq="3\@ym26
\mathchardef\varsupsetneqq="3\@ym27
\mathchardef\subsetneq="3\@ym28
\mathchardef\supsetneq="3\@ym29
\mathchardef\nsubseteq="3\@ym2A
\mathchardef\nsupseteq="3\@ym2B
\mathchardef\nparallel="3\@ym2C
\mathchardef\nmid="3\@ym2D
\mathchardef\nshortmid="3\@ym2E
\mathchardef\nshortparallel="3\@ym2F
\mathchardef\nvdash="3\@ym30
\mathchardef\nVdash="3\@ym31
\mathchardef\nvDash="3\@ym32
\mathchardef\nVDash="3\@ym33
\mathchardef\ntrianglerighteq="3\@ym34
\mathchardef\ntrianglelefteq="3\@ym35
\mathchardef\ntriangleleft="3\@ym36
\mathchardef\ntriangleright="3\@ym37
\mathchardef\nleftarrow="3\@ym38
\mathchardef\nrightarrow="3\@ym39
\mathchardef\nLeftarrow="3\@ym3A
\mathchardef\nRightarrow="3\@ym3B
\mathchardef\nLeftrightarrow="3\@ym3C
\mathchardef\nleftrightarrow="3\@ym3D
\mathchardef\divideontimes="2\@ym3E
\mathchardef\varnothing="0\@ym3F
\mathchardef\nexists="0\@ym40
\mathchardef\mho="0\@ym66
\mathchardef\eth="0\@ym67
\mathchardef\eqsim="3\@ym68
\mathchardef\beth="0\@ym69
\mathchardef\gimel="0\@ym6A
\mathchardef\daleth="0\@ym6B
\mathchardef\lessdot="3\@ym6C
\mathchardef\gtrdot="3\@ym6D
\mathchardef\ltimes="2\@ym6E
\mathchardef\rtimes="2\@ym6F
\mathchardef\shortmid="3\@ym70
\mathchardef\shortparallel="3\@ym71
\mathchardef\smallsetminus="2\@ym72
\mathchardef\thicksim="3\@ym73
\mathchardef\thickapprox="3\@ym74
\mathchardef\approxeq="3\@ym75
\mathchardef\succapprox="3\@ym76
\mathchardef\precapprox="3\@ym77
\mathchardef\curvearrowleft="3\@ym78
\mathchardef\curvearrowright="3\@ym79
\mathchardef\digamma="0\@ym7A
\mathchardef\varkappa="0\@ym7B
\mathchardef\hslash="0\@ym7D
\mathchardef\hbar="0\@ym7E
\mathchardef\backepsilon="3\@ym7F


\def\Bbb{\ifmmode\let\next\Bbb@\else
\def\next{\errmessage{Use \string\Bbb\space only in math mode}}\fi\next}
\def\Bbb@#1{{\Bbb@@{#1}}}
\def\Bbb@@#1{\fam\ymfam#1}
\fi


\def\Nulle{0} 
\def\Afe{1}   
\def\Hae{2}   
\def\Hbe{3}   
\def\Hce{4}   
\def\Hde{5}   


\newcount\LastMac       \LastMac=\Nulle

\newskip\half      \half=5.5pt plus 1.5pt minus 2.25pt
\newskip\one       \one=11pt plus 3pt minus 5.5pt
\newskip\onehalf   \onehalf=16.5pt plus 5.5pt minus 8.25pt
\newskip\two       \two=22pt plus 5.5pt minus 11pt

\def\Half{\addvspace{\half}}
\def\One{\addvspace{\one}}
\def\OneHalf{\addvspace{\onehalf}}
\def\Two{\addvspace{\two}}


\def\Raggedright{
  \rightskip=\z@ plus \hsize\relax
}

\def\Fullout{
  \rightskip=\z@\relax
}

\def\Hang#1#2{
  \hangindent=#1%
  \hangafter=#2\relax
}


\newif\ifsp@page
\def\pagestyle#1{\csname ps@#1\endcsname}
\def\thispagestyle#1{\global\sp@pagetrue\gdef\sp@type{#1}}

\def\ps@titlepage{%
  \def\@oddhead{\eightpoint\noindent \the\CatchLine
    \ifprod@font\else\qquad Printed\ \today\fi \hfil}%
  \let\@evenhead=\@oddhead
}

\def\ps@headings{%
  \def\@oddhead{\elevenpoint\it\noindent
    \hfill\the\RightHeader\hskip1.5em\rm\folio}%
  \def\@evenhead{\elevenpoint\noindent
    \folio\hskip1.5em\it\the\LeftHeader\hfill}%
}

\def\ps@plate{%
  \def\@oddhead{\eightpoint\noindent\plt@cap\hfil}%
  \def\@evenhead{\eightpoint\noindent\plt@cap\hfil}%
}



\def\title#1{
  \bgroup
    \vbox to 8pt{\vss}%
    \seventeenpoint
    \Raggedright
    \noindent \strut{\bf #1}\par
  \egroup
}

\def\author#1{
  \bgroup
    \ifnum\LastMac=\Afe \OneHalf\else \vskip 21pt\fi
    \fourteenpoint
    \Raggedright
    \noindent \strut #1\par
    \vskip 3pt%
  \egroup
}

\def\affiliation#1{
  \bgroup
    \vskip -4pt%
    \eightpoint
    \Raggedright
    \noindent \strut {\it #1}\par
  \egroup
  \LastMac=\Afe\relax
}

\def\acceptedline#1{
  \bgroup
    \Two
    \eightpoint
    \Raggedright
    \noindent \strut #1\par
  \egroup
}

\long\def\abstract#1{%
  \bgroup
    \vskip 20pt%
    \everypar{\Hang{11pc}{0}}%
    \noindent{\ninebf ABSTRACT}\par
    \tenpoint
    \Fullout
    \noindent #1\par
  \egroup
}

\long\def\keywords#1{
  \bgroup
    \Half
    \everypar{\Hang{11pc}{0}}%
    \tenpoint
    \Fullout
    \noindent\hbox{\bf Key words:}\ #1\par
  \egroup
}


\def\maketitle{%
  \EndOpening
  \ifsinglecol \else \MakePage\fi
}


\def\pageoffset#1#2{\hoffset=#1\relax\voffset=#2\relax}


\def\Autonumber{
  \global\AutoNumbertrue  
}

\newif\ifAutoNumber \AutoNumberfalse
\newcount\Sec        
\newcount\SecSec
\newcount\SecSecSec

\Sec=\z@

\def\:{\let\@sptoken= } \:  
\def\:{\@xifnch} \expandafter\def\: {\futurelet\@tempc\@ifnch}

\def\@ifnextchar#1#2#3{%
  \let\@tempMACe #1%
  \def\@tempMACa{#2}%
  \def\@tempMACb{#3}%
  \futurelet \@tempMACc\@ifnch%
}

\def\@ifnch{%
\ifx \@tempMACc \@sptoken%
  \let\@tempMACd\@xifnch%
\else%
  \ifx \@tempMACc \@tempMACe%
    \let\@tempMACd\@tempMACa%
  \else%
    \let\@tempMACd\@tempMACb%
  \fi%
\fi%
\@tempMACd%
}

\def\@ifstar#1#2{\@ifnextchar *{\def\@tempMACa*{#1}\@tempMACa}{#2}}

\newskip\@tempskipb

\def\addvspace#1{%
  \ifvmode\else \endgraf\fi%
  \ifdim\lastskip=\z@%
    \vskip #1\relax%
  \else%
    \@tempskipb#1\relax\@xaddvskip%
  \fi%
}

\def\@xaddvskip{%
  \ifdim\lastskip<\@tempskipb%
    \vskip-\lastskip%
    \vskip\@tempskipb\relax%
  \else%
    \ifdim\@tempskipb<\z@%
      \ifdim\lastskip<\z@ \else%
        \advance\@tempskipb\lastskip%
        \vskip-\lastskip\vskip\@tempskipb%
      \fi%
    \fi%
  \fi%
}

\newskip\@tmpSKIP

\def\addpen#1{%
  \ifvmode
    \if@nobreak
    \else
      \ifdim\lastskip=\z@
        \penalty#1\relax
      \else
        \@tmpSKIP=\lastskip
        \vskip -\lastskip
        \penalty#1\vskip\@tmpSKIP
      \fi
    \fi
  \fi
}

\newcount\@clubpen   \@clubpen=\clubpenalty
\newif\if@nobreak    \@nobreakfalse

\def\@noafterindent{%
  \global\@nobreaktrue
  \everypar{\if@nobreak
              \global\@nobreakfalse
              \clubpenalty \@M
              {\setbox\z@\lastbox}%
              \LastMac=\Nulle\relax%
            \else
              \clubpenalty \@clubpen
              \everypar{}%
            \fi}
}

\newcount\gds@cbrk   \gds@cbrk=-300

\def\@nohdbrk{\interlinepenalty \@M\relax}

\let\@par=\par
\def\@restorepar{\def\par{\@par}}

\newif\if@endpe   \@endpefalse
 
\def\@doendpe{\@endpetrue \@nobreakfalse \LastMac=\Nulle\relax%
     \def\par{\@restorepar\everypar{}\par\@endpefalse}%
              \everypar{\setbox\z@\lastbox\everypar{}\@endpefalse}%
}

\def\section{\@ifstar{\@ssection}{\@section}}

\def\@section#1{
  \if@nobreak
    \everypar{}%
    \ifnum\LastMac=\Hae \addvspace{\half}\fi
  \else
    \addpen{\gds@cbrk}%
    \addvspace{\two}%
  \fi
  \bgroup
    \ninepoint\bf
    \Raggedright
    \ifAutoNumber
      \global\advance\Sec \@ne
      \noindent\@nohdbrk\number\Sec\hskip 1pc \uppercase{#1}\par
      \global\SecSec=\z@
    \else
      \noindent\@nohdbrk\uppercase{#1}\par
    \fi
  \egroup
  \nobreak
  \vskip\half
  \nobreak
  \@noafterindent
  \LastMac=\Hae\relax
}

\def\@ssection#1{
  \if@nobreak
    \everypar{}%
    \ifnum\LastMac=\Hae \addvspace{\half}\fi
  \else
    \addpen{\gds@cbrk}%
    \addvspace{\two}%
  \fi
  \bgroup
    \ninepoint\bf
    \Raggedright
    \noindent\@nohdbrk\uppercase{#1}\par
  \egroup
  \nobreak
  \vskip\half
  \nobreak
  \@noafterindent
  \LastMac=\Hae\relax
}

\def\subsection#1{
  \if@nobreak
    \everypar{}%
    \ifnum\LastMac=\Hae \addvspace{1pt plus 1pt minus .5pt}\fi
  \else
    \addpen{\gds@cbrk}%
    \addvspace{\onehalf}%
  \fi
  \bgroup
    \ninepoint\bf
    \Raggedright
    \ifAutoNumber
      \global\advance\SecSec \@ne
      \noindent\@nohdbrk\number\Sec.\number\SecSec \hskip 1pc\relax #1\par
      \global\SecSecSec=\z@
    \else
      \noindent\@nohdbrk #1\par
    \fi
  \egroup
  \nobreak
  \vskip\half
  \nobreak
  \@noafterindent
  \LastMac=\Hbe\relax
}

\def\subsubsection#1{
  \if@nobreak
    \everypar{}%
    \ifnum\LastMac=\Hbe \addvspace{1pt plus 1pt minus .5pt}\fi
  \else
    \addpen{\gds@cbrk}%
    \addvspace{\onehalf}%
  \fi
  \bgroup
    \ninepoint\it
    \Raggedright
    \ifAutoNumber
      \global\advance\SecSecSec \@ne
      \noindent\@nohdbrk\number\Sec.\number\SecSec.\number\SecSecSec
        \hskip 1pc\relax #1\par
    \else
      \noindent\@nohdbrk #1\par
    \fi
  \egroup
  \nobreak
  \vskip\half
  \nobreak
  \@noafterindent
  \LastMac=\Hce\relax
}

\def\paragraph#1{
  \if@nobreak
    \everypar{}%
  \else
    \addpen{\gds@cbrk}%
    \addvspace{\one}%
  \fi%
  \bgroup%
    \ninepoint\it
    \noindent #1\ \nobreak%
  \egroup
  \LastMac=\Hde\relax
  \ignorespaces
}




\def\beginlist{%
  \par\if@nobreak \else\addvspace{\half}\fi%
  \bgroup%
    \ninepoint
    \let\item=\list@item%
}

\def\list@item{%
  \par\noindent\hskip 1em\relax%
  \ignorespaces%
}

\def\endlist{\par\egroup\addvspace{\half}\@doendpe}


\def\beginrefs{%
  \par
  \bgroup
    \eightpoint
    \Raggedright
    \let\bibitem=\bib@item
}

\def\bib@item{%
  \par\parindent=1.5em\Hang{1.5em}{1}%
  \everypar={\Hang{1.5em}{1}\ignorespaces}%
  \noindent\ignorespaces
}

\def\endrefs{\par\egroup\@doendpe}


\newtoks\CatchLine

\def\@journal{Mon.\ Not.\ R.\ Astron.\ Soc.\ }  
\def\@pubyear{1994}        
\def\@pagerange{000--000}  
\def\@volume{000}          
\def\@microfiche{}         %

\def\pubyear#1{\gdef\@pubyear{#1}\@makecatchline}
\def\pagerange#1{\gdef\@pagerange{#1}\@makecatchline}
\def\volume#1{\gdef\@volume{#1}\@makecatchline}
\def\microfiche#1{\gdef\@microfiche{and Microfiche\ #1}\@makecatchline}

\def\@makecatchline{%
  \global\CatchLine{%
    {\rm \@journal {\bf \@volume},\ \@pagerange\ (\@pubyear)\ \@microfiche}}%
}

\@makecatchline 

\newtoks\LeftHeader
\def\shortauthor#1{
  \global\LeftHeader{#1}%
}

\newtoks\RightHeader
\def\shorttitle#1{
  \global\RightHeader{#1}%
}

\def\PageHead{
  \begingroup
    \ifsp@page
      \csname ps@\sp@type\endcsname
      \global\sp@pagefalse
    \fi
    \ifodd\pageno
      \let\the@head=\@oddhead
    \else
      \let\the@head=\@evenhead
    \fi
    \vbox to \z@{\vskip-22.5\p@%
      \hbox to \PageWidth{\vbox to8.5\p@{}%
        \the@head
      }%
    \vss}%
  \endgroup
  \nointerlineskip
}

\def\today{%
  \number\day\space
  \ifcase\month\or January\or February\or March\or April\or May\or June\or
    July\or August\or September\or October\or November\or December\fi
  \space\number\year%
}

\def\PageFoot{} 

\def\authorcomment#1{%
  \gdef\PageFoot{%
    \nointerlineskip%
    \vbox to 22pt{\vfil%
      \hbox to \PageWidth{\elevenpoint\noindent \hfil #1 \hfil}}%
  }%
}


\newif\ifplate@page
\newbox\plt@box

\def\beginplatepage{%
  \let\plate=\plate@head
  \let\caption=\fig@caption
  \global\setbox\plt@box=\vbox\bgroup
  \TEMPDIMEN=\PageWidth 
  \hsize=\PageWidth\relax
}

\def\endplatepage{\par\egroup\global\plate@pagetrue}
\def\plate@head#1{\gdef\plt@cap{#1}}


\def\letters{%
  \gdef\folio{\ifnum\pageno<\z@ L\romannumeral-\pageno
    \else L\number\pageno \fi}%
}


\everydisplay{\displaysetup}

\newif\ifeqno
\newif\ifleqno

\def\displaysetup#1$${%
 \displaytest#1\eqno\eqno\displaytest
}

\def\displaytest#1\eqno#2\eqno#3\displaytest{%
 \if!#3!\ldisplaytest#1\leqno\leqno\ldisplaytest
 \else\eqnotrue\leqnofalse\def\eqn{#2}\def\eq{#1}\fi
 \generaldisplay$$}

\def\ldisplaytest#1\leqno#2\leqno#3\ldisplaytest{%
 \def\eq{#1}%
 \if!#3!\eqnofalse\else\eqnotrue\leqnotrue
  \def\eqn{#2}\fi}

\def\generaldisplay{%
\ifeqno \ifleqno 
   \hbox to \hsize{\noindent
     $\displaystyle\eq$\hfil$\displaystyle\eqn$}
  \else
    \hbox to \hsize{\noindent
     $\displaystyle\eq$\hfil$\displaystyle\eqn$}
  \fi
 \else
 \hbox to \hsize{\vbox{\noindent
  $\displaystyle\eq$\hfil}}
 \fi
}


\def\@notice{%
  \par\Two%
  \noindent{\b@ls{11pt}\ninerm This paper has been produced using the
    Blackwell Scientific Publications \TeX\ macros.\par}%
}

\outer\def\bye{\@notice\par\vfill\supereject\end}


\def\start@mess{%
  Monthly notices of the RAS journal style (\@typeface)\space
    v\@version,\space \@verdate.%
}

\everyjob{\Warn{\start@mess}}



\newif\if@debug \@debugfalse  

\def\Print#1{\if@debug\immediate\write16{#1}\else \fi}
\def\Warn#1{\immediate\write16{#1}}
\def\wlog#1{}

\newcount\Iteration 

\def\Single{0} \def\Double{1}                 
\def\Figure{0} \def\Table{1}                  

\def\InStack{0}  
\def\InZoneA{1}
\def\InZoneB{2}
\def\InZoneC{3}

\newcount\TEMPCOUNT 
\newdimen\TEMPDIMEN 
\newbox\TEMPBOX     
\newbox\VOIDBOX     

\newcount\LengthOfStack 
\newcount\MaxItems      
\newcount\StackPointer
\newcount\Point         
\newcount\NextFigure    
\newcount\NextTable     
\newcount\NextItem      

\newcount\StatusStack   
\newcount\NumStack      
\newcount\TypeStack     
\newcount\SpanStack     
\newcount\BoxStack      

\newcount\ItemSTATUS    
\newcount\ItemNUMBER    
\newcount\ItemTYPE      
\newcount\ItemSPAN      
\newbox\ItemBOX         
\newdimen\ItemSIZE      

\newdimen\PageHeight    
\newdimen\TextLeading   
\newdimen\Feathering    
\newcount\LinesPerPage  
\newdimen\ColumnWidth   
\newdimen\ColumnGap     
\newdimen\PageWidth     
\newdimen\BodgeHeight   
\newcount\Leading       

\newdimen\ZoneBSize  
\newdimen\TextSize   
\newbox\ZoneABOX     
\newbox\ZoneBBOX     
\newbox\ZoneCBOX     

\newif\ifFirstSingleItem
\newif\ifFirstZoneA
\newif\ifMakePageInComplete
\newif\ifMoreFigures \MoreFiguresfalse 
\newif\ifMoreTables  \MoreTablesfalse  

\newif\ifFigInZoneB 
\newif\ifFigInZoneC 
\newif\ifTabInZoneB 
\newif\ifTabInZoneC

\newif\ifZoneAFullPage

\newbox\MidBOX    
\newbox\LeftBOX
\newbox\RightBOX
\newbox\PageBOX   

\newif\ifLeftCOL  
\LeftCOLtrue

\newdimen\ZoneBAdjust

\newcount\ItemFits
\def\Yes{1}
\def\No{2}


\MaxItems=15
\NextFigure=\z@        
\NextTable=\@ne

\BodgeHeight=6pt
\TextLeading=11pt    
\Leading=11
\Feathering=\z@      
\LinesPerPage=61     
\topskip=\TextLeading
\ColumnWidth=20pc    
\ColumnGap=2pc       

\newskip\ItemSepamount  
\ItemSepamount=\TextLeading plus \TextLeading minus 4pt

\parskip=\z@ plus .1pt
\parindent=18pt
\widowpenalty=\z@
\clubpenalty=10000
\tolerance=1500
\hbadness=1500
\abovedisplayskip=6pt plus 2pt minus 2pt
\belowdisplayskip=6pt plus 2pt minus 2pt
\abovedisplayshortskip=6pt plus 2pt minus 2pt
\belowdisplayshortskip=6pt plus 2pt minus 2pt

\ninepoint 


\PageHeight=682pt

\PageWidth=2\ColumnWidth
\advance\PageWidth by \ColumnGap

\pagestyle{headings}




\newcount\DUMMY \StatusStack=\allocationnumber
\newcount\DUMMY \newcount\DUMMY \newcount\DUMMY 
\newcount\DUMMY \newcount\DUMMY \newcount\DUMMY 
\newcount\DUMMY \newcount\DUMMY \newcount\DUMMY
\newcount\DUMMY \newcount\DUMMY \newcount\DUMMY 
\newcount\DUMMY \newcount\DUMMY \newcount\DUMMY

\newcount\DUMMY \NumStack=\allocationnumber
\newcount\DUMMY \newcount\DUMMY \newcount\DUMMY 
\newcount\DUMMY \newcount\DUMMY \newcount\DUMMY 
\newcount\DUMMY \newcount\DUMMY \newcount\DUMMY 
\newcount\DUMMY \newcount\DUMMY \newcount\DUMMY 
\newcount\DUMMY \newcount\DUMMY \newcount\DUMMY

\newcount\DUMMY \TypeStack=\allocationnumber
\newcount\DUMMY \newcount\DUMMY \newcount\DUMMY 
\newcount\DUMMY \newcount\DUMMY \newcount\DUMMY 
\newcount\DUMMY \newcount\DUMMY \newcount\DUMMY 
\newcount\DUMMY \newcount\DUMMY \newcount\DUMMY 
\newcount\DUMMY \newcount\DUMMY \newcount\DUMMY

\newcount\DUMMY \SpanStack=\allocationnumber
\newcount\DUMMY \newcount\DUMMY \newcount\DUMMY 
\newcount\DUMMY \newcount\DUMMY \newcount\DUMMY 
\newcount\DUMMY \newcount\DUMMY \newcount\DUMMY 
\newcount\DUMMY \newcount\DUMMY \newcount\DUMMY 
\newcount\DUMMY \newcount\DUMMY \newcount\DUMMY

\newbox\DUMMY   \BoxStack=\allocationnumber
\newbox\DUMMY   \newbox\DUMMY \newbox\DUMMY 
\newbox\DUMMY   \newbox\DUMMY \newbox\DUMMY 
\newbox\DUMMY   \newbox\DUMMY \newbox\DUMMY 
\newbox\DUMMY   \newbox\DUMMY \newbox\DUMMY 
\newbox\DUMMY   \newbox\DUMMY \newbox\DUMMY

\def\wlog{\immediate\write\m@ne}


\def\GetItemAll#1{%
 \GetItemSTATUS{#1}
 \GetItemNUMBER{#1}
 \GetItemTYPE{#1}
 \GetItemSPAN{#1}
 \GetItemBOX{#1}
}

\def\GetItemSTATUS#1{%
 \Point=\StatusStack
 \advance\Point by #1
 \global\ItemSTATUS=\count\Point
}

\def\GetItemNUMBER#1{%
 \Point=\NumStack
 \advance\Point by #1
 \global\ItemNUMBER=\count\Point
}

\def\GetItemTYPE#1{%
 \Point=\TypeStack
 \advance\Point by #1
 \global\ItemTYPE=\count\Point
}

\def\GetItemSPAN#1{%
 \Point\SpanStack
 \advance\Point by #1
 \global\ItemSPAN=\count\Point
}

\def\GetItemBOX#1{%
 \Point=\BoxStack
 \advance\Point by #1
 \global\setbox\ItemBOX=\vbox{\copy\Point}
 \global\ItemSIZE=\ht\ItemBOX
 \global\advance\ItemSIZE by \dp\ItemBOX
 \TEMPCOUNT=\ItemSIZE
 \divide\TEMPCOUNT by \Leading
 \divide\TEMPCOUNT by 65536
 \advance\TEMPCOUNT \@ne
 \ItemSIZE=\TEMPCOUNT pt
 \global\multiply\ItemSIZE by \Leading
}


\def\JoinStack{%
 \ifnum\LengthOfStack=\MaxItems 
  \Warn{WARNING: Stack is full...some items will be lost!}
 \else
  \Point=\StatusStack
  \advance\Point by \LengthOfStack
  \global\count\Point=\ItemSTATUS
  \Point=\NumStack
  \advance\Point by \LengthOfStack
  \global\count\Point=\ItemNUMBER
  \Point=\TypeStack
  \advance\Point by \LengthOfStack
  \global\count\Point=\ItemTYPE
  \Point\SpanStack
  \advance\Point by \LengthOfStack
  \global\count\Point=\ItemSPAN
  \Point=\BoxStack
  \advance\Point by \LengthOfStack
  \global\setbox\Point=\vbox{\copy\ItemBOX}
  \global\advance\LengthOfStack \@ne
  \ifnum\ItemTYPE=\Figure 
   \global\MoreFigurestrue
  \else
   \global\MoreTablestrue
  \fi
 \fi
}


\def\LeaveStack#1{%
 {\Iteration=#1
 \loop
 \ifnum\Iteration<\LengthOfStack
  \advance\Iteration \@ne
  \GetItemSTATUS{\Iteration}
   \advance\Point by \m@ne
   \global\count\Point=\ItemSTATUS
  \GetItemNUMBER{\Iteration}
   \advance\Point by \m@ne
   \global\count\Point=\ItemNUMBER
  \GetItemTYPE{\Iteration}
   \advance\Point by \m@ne
   \global\count\Point=\ItemTYPE
  \GetItemSPAN{\Iteration}
   \advance\Point by \m@ne
   \global\count\Point=\ItemSPAN
  \GetItemBOX{\Iteration}
   \advance\Point by \m@ne
   \global\setbox\Point=\vbox{\copy\ItemBOX}
 \repeat}
 \global\advance\LengthOfStack by \m@ne
}


\newif\ifStackNotClean

\def\CleanStack{%
 \StackNotCleantrue
 {\Iteration=\z@
  \loop
   \ifStackNotClean
    \GetItemSTATUS{\Iteration}
    \ifnum\ItemSTATUS=\InStack
     \advance\Iteration \@ne
     \else
      \LeaveStack{\Iteration}
    \fi
   \ifnum\LengthOfStack<\Iteration
    \StackNotCleanfalse
   \fi
 \repeat}
}


\def\FindItem#1#2{%
 \global\StackPointer=\m@ne 
 {\Iteration=\z@
  \loop
  \ifnum\Iteration<\LengthOfStack
   \GetItemSTATUS{\Iteration}
   \ifnum\ItemSTATUS=\InStack
    \GetItemTYPE{\Iteration}
    \ifnum\ItemTYPE=#1
     \GetItemNUMBER{\Iteration}
     \ifnum\ItemNUMBER=#2
      \global\StackPointer=\Iteration
      \Iteration=\LengthOfStack 
     \fi
    \fi
   \fi
  \advance\Iteration \@ne
 \repeat}
}


\def\FindNext{%
 \global\StackPointer=\m@ne 
 {\Iteration=\z@
  \loop
  \ifnum\Iteration<\LengthOfStack
   \GetItemSTATUS{\Iteration}
   \ifnum\ItemSTATUS=\InStack
    \GetItemTYPE{\Iteration}
   \ifnum\ItemTYPE=\Figure
    \ifMoreFigures
      \global\NextItem=\Figure
      \global\StackPointer=\Iteration
      \Iteration=\LengthOfStack 
    \fi
   \fi
   \ifnum\ItemTYPE=\Table
    \ifMoreTables
      \global\NextItem=\Table
      \global\StackPointer=\Iteration
      \Iteration=\LengthOfStack 
    \fi
   \fi
  \fi
  \advance\Iteration \@ne
 \repeat}
}


\def\ChangeStatus#1#2{%
 \Point=\StatusStack
 \advance\Point by #1
 \global\count\Point=#2
}



\def\Zone{\InZoneA}

\ZoneBAdjust=\z@

\def\MakePage{
 \global\ZoneBSize=\PageHeight
 \global\TextSize=\ZoneBSize
 \global\ZoneAFullPagefalse
 \global\topskip=\TextLeading
 \MakePageInCompletetrue
 \MoreFigurestrue
 \MoreTablestrue
 \FigInZoneBfalse
 \FigInZoneCfalse
 \TabInZoneBfalse
 \TabInZoneCfalse
 \global\FirstSingleItemtrue
 \global\FirstZoneAtrue
 \global\setbox\ZoneABOX=\box\VOIDBOX
 \global\setbox\ZoneBBOX=\box\VOIDBOX
 \global\setbox\ZoneCBOX=\box\VOIDBOX
 \loop
  \ifMakePageInComplete
 \FindNext
 \ifnum\StackPointer=\m@ne
  \NextItem=\m@ne
  \MoreFiguresfalse
  \MoreTablesfalse
 \fi
 \ifnum\NextItem=\Figure
   \FindItem{\Figure}{\NextFigure}
   \ifnum\StackPointer=\m@ne \global\MoreFiguresfalse
   \else
    \GetItemSPAN{\StackPointer}
    \ifnum\ItemSPAN=\Single \def\Zone{\InZoneB}\relax
     \ifFigInZoneC \global\MoreFiguresfalse\fi
    \else
     \def\Zone{\InZoneA}
     \ifFigInZoneB \def\Zone{\InZoneC}\fi
    \fi
   \fi
   \ifMoreFigures\Print{}\FigureItems\fi
 \fi
\ifnum\NextItem=\Table
   \FindItem{\Table}{\NextTable}
   \ifnum\StackPointer=\m@ne \global\MoreTablesfalse
   \else
    \GetItemSPAN{\StackPointer}
    \ifnum\ItemSPAN=\Single\relax
     \ifTabInZoneC \global\MoreTablesfalse\fi
    \else
     \def\Zone{\InZoneA}
     \ifTabInZoneB \def\Zone{\InZoneC}\fi
    \fi
   \fi
   \ifMoreTables\Print{}\TableItems\fi
 \fi
   \MakePageInCompletefalse 
   \ifMoreFigures\MakePageInCompletetrue\fi
   \ifMoreTables\MakePageInCompletetrue\fi
 \repeat
 \ifZoneAFullPage
  \global\TextSize=\z@
  \global\ZoneBSize=\z@
  \global\vsize=\z@\relax
  \global\topskip=\z@\relax
  \vbox to \z@{\vss}
  \eject
 \else
 \global\advance\ZoneBSize by -\ZoneBAdjust
 \global\vsize=\ZoneBSize
 \global\hsize=\ColumnWidth
 \global\ZoneBAdjust=\z@
 \ifdim\TextSize<23pt
 \Warn{}
 \Warn{* Making column fall short: TextSize=\the\TextSize *}
 \vskip-\lastskip\eject\fi
 \fi
}

\def\MakeRightCol{
 \global\TextSize=\ZoneBSize
 \MakePageInCompletetrue
 \MoreFigurestrue
 \MoreTablestrue
 \global\FirstSingleItemtrue
 \global\setbox\ZoneBBOX=\box\VOIDBOX
 \def\Zone{\InZoneB}
 \loop
  \ifMakePageInComplete
 \FindNext
 \ifnum\StackPointer=\m@ne
  \NextItem=\m@ne
  \MoreFiguresfalse
  \MoreTablesfalse
 \fi
 \ifnum\NextItem=\Figure
   \FindItem{\Figure}{\NextFigure}
   \ifnum\StackPointer=\m@ne \MoreFiguresfalse
   \else
    \GetItemSPAN{\StackPointer}
    \ifnum\ItemSPAN=\Double\relax
     \MoreFiguresfalse\fi
   \fi
   \ifMoreFigures\Print{}\FigureItems\fi
 \fi
 \ifnum\NextItem=\Table
   \FindItem{\Table}{\NextTable}
   \ifnum\StackPointer=\m@ne \MoreTablesfalse
   \else
    \GetItemSPAN{\StackPointer}
    \ifnum\ItemSPAN=\Double\relax
     \MoreTablesfalse\fi
   \fi
   \ifMoreTables\Print{}\TableItems\fi
 \fi
   \MakePageInCompletefalse 
   \ifMoreFigures\MakePageInCompletetrue\fi
   \ifMoreTables\MakePageInCompletetrue\fi
 \repeat
 \ifZoneAFullPage
  \global\TextSize=\z@
  \global\ZoneBSize=\z@
  \global\vsize=\z@\relax
  \global\topskip=\z@\relax
  \vbox to \z@{\vss}
  \eject
 \else
 \global\vsize=\ZoneBSize
 \global\hsize=\ColumnWidth
 \ifdim\TextSize<23pt
 \Warn{}
 \Warn{* Making column fall short: TextSize=\the\TextSize *}
 \vskip-\lastskip\eject\fi
\fi
}

\def\FigureItems{
 \Print{Considering...}
 \ShowItem{\StackPointer}
 \GetItemBOX{\StackPointer} 
 \GetItemSPAN{\StackPointer}
  \CheckFitInZone 
  \ifnum\ItemFits=\Yes
   \ifnum\ItemSPAN=\Single
     \ChangeStatus{\StackPointer}{\InZoneB} 
     \global\FigInZoneBtrue
     \ifFirstSingleItem
      \hbox{}\vskip-\BodgeHeight
     \global\advance\ItemSIZE by \TextLeading
     \fi
     \unvbox\ItemBOX\ItemSep
     \global\FirstSingleItemfalse
     \global\advance\TextSize by -\ItemSIZE
     \global\advance\TextSize by -\TextLeading
   \else
    \ifFirstZoneA
     \global\advance\ItemSIZE by \TextLeading
     \global\FirstZoneAfalse\fi
    \global\advance\TextSize by -\ItemSIZE
    \global\advance\TextSize by -\TextLeading
    \global\advance\ZoneBSize by -\ItemSIZE
    \global\advance\ZoneBSize by -\TextLeading
    \ifFigInZoneB\relax
     \else
     \ifdim\TextSize<3\TextLeading
     \global\ZoneAFullPagetrue
     \fi
    \fi
    \ChangeStatus{\StackPointer}{\Zone}
    \ifnum\Zone=\InZoneC \global\FigInZoneCtrue\fi
  \fi
   \Print{TextSize=\the\TextSize}
   \Print{ZoneBSize=\the\ZoneBSize}
  \global\advance\NextFigure \@ne
   \Print{This figure has been placed.}
  \else
   \Print{No space available for this figure...holding over.}
   \Print{}
   \global\MoreFiguresfalse
  \fi
}

\def\TableItems{
 \Print{Considering...}
 \ShowItem{\StackPointer}
 \GetItemBOX{\StackPointer} 
 \GetItemSPAN{\StackPointer}
  \CheckFitInZone 
  \ifnum\ItemFits=\Yes
   \ifnum\ItemSPAN=\Single
    \ChangeStatus{\StackPointer}{\InZoneB}
     \global\TabInZoneBtrue
     \ifFirstSingleItem
      \hbox{}\vskip-\BodgeHeight
     \global\advance\ItemSIZE by \TextLeading
     \fi
     \unvbox\ItemBOX\ItemSep
     \global\FirstSingleItemfalse
     \global\advance\TextSize by -\ItemSIZE
     \global\advance\TextSize by -\TextLeading
   \else
    \ifFirstZoneA
    \global\advance\ItemSIZE by \TextLeading
    \global\FirstZoneAfalse\fi
    \global\advance\TextSize by -\ItemSIZE
    \global\advance\TextSize by -\TextLeading
    \global\advance\ZoneBSize by -\ItemSIZE
    \global\advance\ZoneBSize by -\TextLeading
    \ifFigInZoneB\relax
     \else
     \ifdim\TextSize<3\TextLeading
     \global\ZoneAFullPagetrue
     \fi
    \fi
    \ChangeStatus{\StackPointer}{\Zone}
    \ifnum\Zone=\InZoneC \global\TabInZoneCtrue\fi
   \fi
  \global\advance\NextTable \@ne
   \Print{This table has been placed.}
  \else
  \Print{No space available for this table...holding over.}
   \Print{}
   \global\MoreTablesfalse
  \fi
}


\def\CheckFitInZone{%
{\advance\TextSize by -\ItemSIZE
 \advance\TextSize by -\TextLeading
 \ifFirstSingleItem
  \advance\TextSize by \TextLeading
 \fi
 \ifnum\Zone=\InZoneA\relax
  \else \advance\TextSize by -\ZoneBAdjust
 \fi
 \ifdim\TextSize<3\TextLeading \global\ItemFits=\No
 \else \global\ItemFits=\Yes\fi}
}

\def\BeginOpening{%
  \thispagestyle{titlepage}%
  \global\setbox\ItemBOX=\vbox\bgroup%
    \hsize=\PageWidth%
    \hrule height \z@
    \ifsinglecol\vskip 6pt\fi 
}

\let\begintopmatter=\BeginOpening  

\def\EndOpening{%
  \One
  \egroup
  \ifsinglecol
    \box\ItemBOX%
    \vskip\TextLeading plus 2\TextLeading
    \@noafterindent
  \else
    \ItemNUMBER=\z@%
    \ItemTYPE=\Figure
    \ItemSPAN=\Double
    \ItemSTATUS=\InStack
    \JoinStack
  \fi
}


\newif\if@here  \@herefalse

\def\no@float{\global\@heretrue}
\let\nofloat=\relax 

\def\beginfigure{%
  \@ifstar{\global\@dfloattrue \@bfigure}{\global\@dfloatfalse \@bfigure}%
}

\def\@bfigure#1{%
  \par
  \if@dfloat
    \ItemSPAN=\Double
    \TEMPDIMEN=\PageWidth
  \else
    \ItemSPAN=\Single
    \TEMPDIMEN=\ColumnWidth
  \fi
  \ifsinglecol
    \TEMPDIMEN=\PageWidth
  \else
    \ItemSTATUS=\InStack
    \ItemNUMBER=#1%
    \ItemTYPE=\Figure
  \fi
  \bgroup
    \hsize=\TEMPDIMEN
    \global\setbox\ItemBOX=\vbox\bgroup
      \eightpoint\nostb@ls{10pt}%
      \let\caption=\fig@caption
      \ifsinglecol \let\nofloat=\no@float\fi
}

\def\fig@caption#1{%
  \vskip 5.5pt plus 6pt%
  \bgroup 
    \eightpoint\nostb@ls{10pt}%
    \setbox\TEMPBOX=\hbox{#1}%
    \ifdim\wd\TEMPBOX>\TEMPDIMEN
      \noindent \unhbox\TEMPBOX\par
    \else
      \hbox to \hsize{\hfil\unhbox\TEMPBOX\hfil}%
    \fi
  \egroup
}

\def\endfigure{%
  \par\egroup 
  \egroup
  \ifsinglecol
    \if@here \midinsert\global\@herefalse\else \topinsert\fi
      \unvbox\ItemBOX
    \endinsert
  \else
    \JoinStack
    \Print{Processing source for figure \the\ItemNUMBER}%
  \fi
}


\newbox\tab@cap@box
\def\tab@caption#1{\global\setbox\tab@cap@box=\hbox{#1\par}}

\newtoks\tab@txt@toks
\long\def\tab@txt#1{\global\tab@txt@toks={#1}\global\table@txttrue}

\newif\iftable@txt  \table@txtfalse
\newif\if@dfloat    \@dfloatfalse

\def\begintable{%
  \@ifstar{\global\@dfloattrue \@btable}{\global\@dfloatfalse \@btable}%
}

\def\@btable#1{%
  \par
  \if@dfloat
    \ItemSPAN=\Double
    \TEMPDIMEN=\PageWidth
  \else
    \ItemSPAN=\Single
    \TEMPDIMEN=\ColumnWidth
  \fi
  \ifsinglecol
    \TEMPDIMEN=\PageWidth
  \else
    \ItemSTATUS=\InStack
    \ItemNUMBER=#1%
    \ItemTYPE=\Table
  \fi
  \bgroup
    \eightpoint\nostb@ls{10pt}%
    \global\setbox\ItemBOX=\vbox\bgroup
      \let\caption=\tab@caption
      \let\tabletext=\tab@txt
      \ifsinglecol \let\nofloat=\no@float\fi
}

\def\endtable{%
  \par\egroup 
  \egroup
  \setbox\TEMPBOX=\hbox to \TEMPDIMEN{%
    \hss
    \vbox{%
      \hsize=\wd\ItemBOX
      \ifvoid\tab@cap@box
      \else
        \noindent\unhbox\tab@cap@box
        \vskip 5.5pt plus 6pt%
      \fi
      \box\ItemBOX
      \iftable@txt
        \vskip 10pt%
        \eightpoint\nostb@ls{10pt}%
        \noindent\the\tab@txt@toks
        \global\table@txtfalse
      \fi
    }%
    \hss
  }%
  \ifsinglecol
    \if@here \midinsert\global\@herefalse\else \topinsert\fi
      \box\TEMPBOX
    \endinsert
  \else
    \global\setbox\ItemBOX=\box\TEMPBOX
    \JoinStack
    \Print{Processing source for table \the\ItemNUMBER}%
  \fi
}

\def\UnloadZoneA{%
\FirstZoneAtrue
 \Iteration=\z@
  \loop
   \ifnum\Iteration<\LengthOfStack
    \GetItemSTATUS{\Iteration}
    \ifnum\ItemSTATUS=\InZoneA
     \GetItemBOX{\Iteration}
     \ifFirstZoneA \vbox to \BodgeHeight{\vfil}%
     \FirstZoneAfalse\fi
     \unvbox\ItemBOX\ItemSep
     \LeaveStack{\Iteration}
     \else
     \advance\Iteration \@ne
   \fi
 \repeat
}

\def\UnloadZoneC{%
\Iteration=\z@
  \loop
   \ifnum\Iteration<\LengthOfStack
    \GetItemSTATUS{\Iteration}
    \ifnum\ItemSTATUS=\InZoneC
     \GetItemBOX{\Iteration}
     \ItemSep\unvbox\ItemBOX
     \LeaveStack{\Iteration}
     \else
     \advance\Iteration \@ne
   \fi
 \repeat
}


\def\ShowItem#1{
  {\GetItemAll{#1}
  \Print{\the#1:
  {TYPE=\ifnum\ItemTYPE=\Figure Figure\else Table\fi}
  {NUMBER=\the\ItemNUMBER}
  {SPAN=\ifnum\ItemSPAN=\Single Single\else Double\fi}
  {SIZE=\the\ItemSIZE}}}
}

\def\ShowStack{%
 \Print{}
 \Print{LengthOfStack = \the\LengthOfStack}
 \ifnum\LengthOfStack=\z@ \Print{Stack is empty}\fi
 \Iteration=\z@
 \loop
 \ifnum\Iteration<\LengthOfStack
  \ShowItem{\Iteration}
  \advance\Iteration \@ne
 \repeat
}

\def\B#1#2{%
\hbox{\vrule\kern-0.4pt\vbox to #2{%
\hrule width #1\vfill\hrule}\kern-0.4pt\vrule}
}


\newif\ifsinglecol   \singlecolfalse

\def\onecolumn{%
  \global\output={\singlecoloutput}%
  \global\hsize=\PageWidth
  \global\vsize=\PageHeight
  \global\ColumnWidth=\hsize
  \global\TextLeading=12pt
  \global\Leading=12
  \global\singlecoltrue
  \global\let\onecolumn=\relax
  \global\let\footnote=\sing@footnote
  \global\let\vfootnote=\sing@vfootnote
  \ninepoint 
  \message{(Single column)}%
}

\def\singlecoloutput{%
  \shipout\vbox{\PageHead\pagebody\PageFoot}%
  \advancepageno
  \ifplate@page
    \shipout\vbox{%
      \sp@pagetrue
      \def\sp@type{plate}%
      \global\plate@pagefalse
      \PageHead\vbox to \PageHeight{\unvbox\plt@box\vfil}\PageFoot%
    }%
    \message{[plate]}%
    \advancepageno
  \fi
  \ifnum\outputpenalty>-\@MM \else\dosupereject\fi%
}

\def\ItemSep{\vskip\ItemSepamount\relax}

\def\ItemSepbreak{\par\ifdim\lastskip<\ItemSepamount
  \removelastskip\penalty-200\ItemSep\fi%
}


\let\@@endinsert=\endinsert 

\def\endinsert{\egroup 
  \if@mid \dimen@\ht\z@ \advance\dimen@\dp\z@ \advance\dimen@12\p@
    \advance\dimen@\pagetotal \advance\dimen@-\pageshrink
    \ifdim\dimen@>\pagegoal\@midfalse\p@gefalse\fi\fi
  \if@mid \ItemSep\box\z@\ItemSepbreak
  \else\insert\topins{\penalty100 
    \splittopskip\z@skip
    \splitmaxdepth\maxdimen \floatingpenalty\z@
    \ifp@ge \dimen@\dp\z@
    \vbox to\vsize{\unvbox\z@\kern-\dimen@}
    \else \box\z@\nobreak\ItemSep\fi}\fi\endgroup%
}


\def\gobbleone#1{}
\def\gobbletwo#1#2{}
\let\footnote=\gobbletwo 
\let\vfootnote=\gobbleone

\def\sing@footnote#1{\let\@sf\empty 
  \ifhmode\edef\@sf{\spacefactor\the\spacefactor}\/\fi
  \hbox{$^{\hbox{\eightpoint #1}}$}\@sf\sing@vfootnote{#1}%
}

\def\sing@vfootnote#1{\insert\footins\bgroup\eightpoint\b@ls{9pt}%
  \interlinepenalty\interfootnotelinepenalty
  \splittopskip\ht\strutbox 
  \splitmaxdepth\dp\strutbox \floatingpenalty\@MM
  \leftskip\z@skip \rightskip\z@skip \spaceskip\z@skip \xspaceskip\z@skip
  \noindent $^{\scriptstyle\hbox{#1}}$\hskip 4pt%
    \footstrut\futurelet\next\fo@t%
}

\def\footnoterule{\kern-3\p@ \hrule height \z@ \kern 3\p@}

\skip\footins=19.5pt plus 12pt minus 1pt
\count\footins=1000
\dimen\footins=\maxdimen


\def\landscape{%
  \global\TEMPDIMEN=\PageWidth
  \global\PageWidth=\PageHeight
  \global\PageHeight=\TEMPDIMEN
  \global\let\landscape=\relax
  \onecolumn
  \message{(landscape)}%
  \raggedbottom
}


\output{%
  \ifLeftCOL
    \global\setbox\LeftBOX=\vbox to \ZoneBSize{\box255\unvbox\ZoneBBOX}%
    \global\LeftCOLfalse
    \MakeRightCol
  \else
    \setbox\RightBOX=\vbox to \ZoneBSize{\box255\unvbox\ZoneBBOX}%
    \setbox\MidBOX=\hbox{\box\LeftBOX\hskip\ColumnGap\box\RightBOX}%
    \setbox\PageBOX=\vbox to \PageHeight{%
      \UnloadZoneA\box\MidBOX\UnloadZoneC}%
    \shipout\vbox{\PageHead\box\PageBOX\PageFoot}%
    \advancepageno
    \ifplate@page
      \shipout\vbox{%
        \sp@pagetrue
        \def\sp@type{plate}%
        \global\plate@pagefalse
        \PageHead\vbox to \PageHeight{\unvbox\plt@box\vfil}\PageFoot%
      }%
      \message{[plate]}%
      \advancepageno
    \fi
    \global\LeftCOLtrue
    \CleanStack
    \MakePage
  \fi
}


\Warn{\start@mess}

\def\mnmacrosloaded{} 

\catcode `\@=12 



 \fi
\loadboldmathnames
\pageoffset{-2.5pc}{0pc}



\Autonumber

\begintopmatter

\title{Models of optical/UV continuum in AGN.\break
Constraints from NGC 5548 monitoring campaign.}
\author{Z. Loska$^{1}$ and  B. Czerny$^{1}$,
 }
\affiliation{$^1$ Nicolaus Copernicus Astronomical Center, Bartycka 18, 
00--716 Warsaw, Poland}

\shortauthor{Z. Loska and B. Czerny}
\shorttitle{Models of optical/UV continuum of NGC 5548}

\abstract{We analyse the data from the
optical/IUE observational campaign of the 
Seyfert galaxy NGC 5548 in the context of 10 phenomenological models.
On the basis of the optical/UV data as well as constraints from the
X-ray observations we can favour one model of the nucleus: an accretion disc
with an inner radius cutoff surrounded by a hot corona. The second acceptable 
model for optical/UV data is a distribution of optically thin clouds. 
However, X-ray
constraints which were crucial in the analysis of disc type models could not
have been applied; further development of this model is necessary.

}

\keywords {accretion, accretion discs -- galaxies: active -- galaxies: Seyfert
-- X-rays: galaxies}

\maketitle

\section{Introduction}

The nearby spiral galaxy NGC 5548 (z= 0.0174) is one of the best 
laboratories for testing AGN models
(Rokaki, Collin-Souffrin \& Magnan 1993). The nucleus is bright (absolute 
visual magnitude -24.3), the interstellar extinction in this direction is
exceptionally low (hydrogen column $N_H \sim 1.65 \times 
10^{20}$, Nandra et al. 1991)
and the nucleus is strongly
variable in optical, UV and X-ray band. Extensive monitoring
of that galaxy is reported in a number of papers devoted to optical
(Peterson et al. 1991, Peterson et al. 1992, Peterson et al. 1994)
UV (Clavel et al. 1991), and simultaneous Ginga \& IUE campaign 
(Clavel et al. 1992). 
It was also one of the
few AGN detected in usually unobserved EUV band (Marshall, Fruscione \&
Carone 1995).

The overall IR-UV spectrum (e.g. Ward et al. 1987) and X-ray spectrum 
(e.g. Nandra \& Pounds 1994, Done et al. 1995) 
of the nucleus of NGC 5548 is fairly typical for a Seyfert 1 galaxy although
the spectroscopic classification identifies it usually as Seyfert 1.5.    

The amplitude of the variability of NGC 5548 is considerable at every wavelength
from optical to hard X-rays.

In the optical band significant
 fraction of the luminosity is thought to be due to the
contribution of starlight (e.g. Kotilainen \& Ward 1994). 
The flux measured  at 5100 \AA~ varied by a factor 2 
during the four years of the campaign (1988 --1992). UV
flux monitored in 1988 -- 1989 varied almost by a factor 3 at 1350 \AA. 

As the continuum variations in all three UV bands and the optical band were
simultaneous (down to a  measurable delay of a few days; Clavel et al. 1991), 
most of the optical/UV variability can be accounted
for by the reprocessing of hard X-rays, as suggested by  Malkan (1991) and
Collin-Souffrin (1991) (see also Krolik et al. 1991, Clavel et al. 1992,
Rokaki et al. 1993). A dominant role for reprocessing is also 
consistent with Ginga
data (Nandra \& Pounds 1994) which clearly show the presence
of the reflection component superimposed on a typical hard X-ray power law
(energy index $\sim 0.9$), with reflected fraction of order 
the usual 2$\pi$.
Recent ASCA data (Mushotzky et al. 1995) support this conclusion
as they show  the $K_{\alpha}$ line profile consistent with a disc 
around a Kerr
black hole inclined $\sim 15^o - 38^o$ with respect to an 
observer. As the iron line is broad (FWHM $>$ 35 000 km/s) most of the
emission comes from the region smaller than $\sim 36$ Schwarzschild radii. 
In most epochs
the X-ray flux above 2 keV is strongly variable (factor 2 between separate
observations, Nandra \& Pounds 1994) 
but without systematic trends in the change of the energy index. 
In soft X-rays the 
variability pattern is more complex due to the presence of soft X-ray excess
and the warm absorber (see Nandra et al. 1993 and Done et al. 1995 for 
ROSAT observations).

However,
occasionally strong enhancements of the big bump component are seen
(in UV in May 1984, Clavel et al. 1992; in soft X-rays in December 
1992/January 1993, Done et al. 
1995) which are unrelated to hard X-ray luminosity and indicate direct 
liberation of gravitational energy in the form of optically thick emission
in these periods. 

Multiwavelength studies put constraints on any models of the nucleus
by  determining or limiting the connections and delays between spectral
bands. As for the continuum, both optical and UV spectra follow X-ray flux
(with exceptions mentioned above) with the delay less than 6 days 
(Clavel et al. 1992).
In the case of UV this delay is most probably smaller than 2 days which 
suggests the size of the UV emitting region of order of $5 \times 10^{15}$
cm. Large amplitude changes
in soft X-ray emission happen in 2 days and large amplitude variations 
of hard X-rays also require time scales of order of two days (e.g. Done et al. 1995).

The available data was mostly used to study the structure of the Broad Line 
Region (e.g. Done \& Krolik 1996) as in that case the 
delays are well measurable.

A model of the continuum emission of NGC 5548 was elaborated by Rokaki
et al. (1993). The model consisted of an inner disc
emitting X-rays (modelled as a sphere of a constant radius and variable 
luminosity), an outer standard disc and an optically thin surrounding medium.
The outer disc was irradiated by X-rays either directly or by photons scattered
by the thin medium. Such a model adequately represented the data from IUE
campaign but formal fits were not presented so a quantitative analysis
was impossible.

In this paper we discuss a number of simple phenomenological models and we
fit them to the data from the same observational campaign. We show that
for the mass of the black hole $6 \times 10^7 M_{\odot}$ (as estimated on
the basis of X-ray reprocessing, UV and X-ray variability and emission lines) 
only two of these models are actually acceptable if the optical/UV data
requirements are supplemented by the constraints from (non-simultaneous)
X-ray data. The first of these two models is a 
non-stationary accretion disc with variable accretion rate and a variable
inner radius cut-off. The gravitational energy available below the
cut-off radius is liberated in the form of X-rays in the hot corona surrounding
the disc and thermalized within the disc. The second model is the
distribution of the (identical) 
optically thin clouds which emission in the optical/UV band can be approximated
as a free-free emission. 

The content of the paper is following. Models are described in Section 2.
In Section 3 we describe the method of analysis of the data including the
problem of the starlight contribution and the contribution of Balmer continuum
and blended FeII lines at 2670 \AA. The results are given in Section 4.
Section 5 contains the discussion of the results. Conclusions are given
in Section 6. 

\section{Models of optical/UV continuum}

We introduce a number of viable models describing the extension of the 
optically thick disc-like part of the flow and the geometry of the source
of X-ray radiation. As there are no detailed theoretical predictions about the 
formation of these two phases of accreting gas we introduce phenomenological
parameters but with a physically sound interpretation. All models considered 
here do not contain more than two free time-dependent parameters as we fit only
a time sequence of four frequency points. However, models may contain
additional parameters not varying with time, like the mass of the central 
black hole etc., as they do not reduce noticeably the number of the degrees of
freedom.

The first three models are based on the assumption that the only variable
component is the X-ray flux incident upon a stationary accretion disc. The 
next five models are based on variations in the accretion rate at the innermost
part of the disc which in turn may cause a change in irradiation. 
As a single optical/UV spectrum consists of only 4 points we approximate the
local thermal emission of the disc by a black body with the effective 
temperature distribution calculated including the viscous dissipation 
(determined by accretion rate) and the irradiation flux appropriate for adopted
geometry (taking albedo equal 0.9). We assume non-rotating black hole, i.e.
Schwarzschild geometry.

The next
model is the simplest version of emission by optically thin clouds.
Finally, we also fit a single power law model for a better discussion of the 
quality of the data.

\subsection{Stationary accretion disc and a point-like X-ray source}

This model consists of a standard stationary Keplerian disc around a massive
black hole of the mass $M$ and accretion rate $\dot M$, 
and a point-like source of X-ray radiation localized on the disc
symmetry axis and characterized by the 
luminosity $L_X$ and distance from the disc plane $H_X$. 
Such a geometry was frequently suggested and it was used in a number of 
papers devoted to modelling AGN 
emission (e.g. Ross \& Fabian 1993, \. Zycki \& Czerny 1994). 
It may for example approximate the situation where
hard X-ray emission is produced by shocks which may accompany the 
formation of the jet (e.g. Henri \&
Pelletier 1991, Liang \& Li 1995). Since it was suggested that the variable 
irradiation is
mostly responsible for the variable optical/UV emission (Clavel et al. 1991,
Rokaki et al. 1993) we assume that
 $\dot M$ is constant whilst $L_X$ and $H_X$ vary. The local 
effective temperature in the disc is calculated by taking into account the
stationary viscous flux and the thermalized X-ray flux. Computing the last
term we assume that the albedo for X-rays is equal 0.9 as determined by
Lightman \& White (1988) (see also \. Zycki et al. 1994) 
as the reflection component observed in hard X-ray
data indicates the presence of neutral gas (Nandra et al. 1991). 

\subsection{Stationary accretion disc with geometrically thick 
optically thin inner part}

As the standard accretion disc model with $\alpha P$ viscosity (Shakura \& 
Sunyaev 1973) is thermally unstable in the inner parts the inner disc may 
become optically thin and hot. Such a flow was studied in a number of papers
(Shapiro, Lightman \& Eardley 1976, Wandel \& Liang 1991, Kusunose \& 
Zdziarski 1994).
The physics behind this is complex but a simplified geometrical model can be used:
the transition radius $R_X$ defines the cut-off radius for the optically thick 
standard disc (e.g. Siemiginowska \& Czerny 1989). The inner part of the disc 
is replaced with inner optically thin disc of the luminosity $L_X$.
The shape of the inner disc is approximated by a sphere of a radius $R_X$. 
We do not require that the gravitational energy available below $R_X$ is equal
to $L_X$ in order to account for possible magnetic storing of the energy
in the optically thin region. The sphere irradiates the outer optically thick
parts of the disc. A similar model was actually
fitted to the light curve of NGC 5548 by Rokaki et al. (1993). The difference
between their model and the one presented here is in their assumption of 
constant $R_X$ ($3 \times 10^{14}$ cm) and the presence of 
additional irradiation of the disc due to an extended hot medium. 
In our model we assume a constant accretion
rate in the outer parts of the disc but allow for variations of both $L_X$ and 
$R_X$. It again implies some storage of the energy by the inner disc.

\subsection{Stationary accretion disc irradiated by a hot corona}

A number of papers were devoted to the scenario in which most of the energy is
dissipated in the disc corona instead of the disc interior (e.g. Liang 
\& Price 1977, Paczy\' nski 1978; \. Zycki, Collin-Souffrin \& Czerny 1995).
We roughly model such a situation assuming that standard stationary accretion disc
of a given $M$ and $\dot M$ is irradiated by the corona effectively
producing the local flux 
$$F_{cor}=A(r/3r_g)^{-\beta}\eqno(1)$$
where both A and the index $\beta$ were allowed to vary. To express the
model parameters more conveniently we use the bolometric luminosity of 
the corona rather than the coefficient A (both are easily related by the integration
over radius, i.e. $L_X= 18 \pi A r_g^2/[(r_{out}/3r_g)^{2-\beta} -1]/(2 - \beta)$  
for $\beta$ different from 2).
In this model the strength of the corona is unrelated to the accretion rate.

\subsection{Non-stationary accretion disc with advection-dominated 
innermost part}

The timescale of the UV variability of order of a few days is actually 
consistent with the viscous timescales of the inner few Schwarzschild radii
if the viscosity parameter is high (see e.g. Siemiginowska \& Czerny 1989).
Therefore we study the hypothesis that the observed spectral changes are
actually driven by the variations in the accretion rate. In this model we
assume that the innermost part of the disc may become optically thin. If such
optically thin flow is dominated by advection (e.g. Narayan \& Yi 1995) 
only a small fraction of energy released in this part of the flow would emerge
in the form of X-rays and irradiate 
the optically thick parts. Since this X-ray emission 
can then be an arbitrarily low fraction
of the available gravitational energy we neglect the irradiation as the data
is not good enough to introduce another free parameter.

Therefore in this model
we simply assume that the standard disc extends down to the  a certain 
cut-off radius $R_C$ and the part of the disc below does not contribute 
to UV radiation flux.

\subsection{Non-stationary accretion disc with inner radius cut-off and a
point-like X-ray source}

If the gravitational energy cannot be stored efficiently by the magnetic field
the model 2.1 of the point-like source is incorrect. Therefore we
consider another model in which an accretion disc extends only down to a certain
cut-off radius $R_C$. The remaining available energy not dissipated 
in the disc
is now dissipated in the point-like source. Such a model is motivated
for example by suggested ejection of plasmoids at the expense of the
gravitational energy of the inner part of the disc (e.g. Liang \& Li
1995). We assume that the X-ray
source is located on the symmetry axis at the distance $R_C$.
We allow now for variable accretion rate in the disc $\dot M$ as the
cut-off radius may (in fact, should) depend on it. Therefore the model
again has two variable parameters.

\subsection{Non-stationary accretion disc with geometrically thick optically 
thin inner part}

If the energy cannot be stored efficiently by the inner hot part of the disc
as it can in model 2.2 the luminosity of this part has to be matched by the 
available gravitational
energy below $R_{X}$. In close analogy with model 2.2 we now choose the
disc accretion rate $\dot M$ and the cut-off radius $R_X$ filled by the
spherical inner disc as our variable parameters; the bolometric
luminosity of the sphere and the effect of irradiation of the disc by its
X-ray emission are uniquely determined by these two values. The model is also 
complementary to model 2.4  as this time we neglect the advection losses and 
we allow for the irradiation of the
optically thick parts of the disc.

\subsection{Non-stationary accretion disc with inner radius cut-off
and scattered X-ray radiation}

This model differs with respect to the previous two in the geometry of the
irradiation by X-ray. This time we assume that the luminosity is released
in the center and redirected towards the disc by scattering in an optically
thin hot medium. Effectively, we just make the assumption that the incident
luminosity is inversely proportional to the square of a distance. Similar
dependence on a distance was adopted in the next model  but this time the
cut-off radius enters the luminosity only through the total X-ray luminosity
whilst in the corona model 2.8 the cut-off radius would enter also through the
normalization.

\subsection{Non-stationary accretion disc with inner radius cut-off irradiated by the corona}

This model differs from the model 2.3  as now we consider the specific
example of the dependence of the corona parameter on the accretion rate
in the disc. We now assume that the standard disc extends not to the
marginally stable orbit but only to a certain cut-off radius $R_C$ as
in model 2.4. The gravitational energy below this radius is now exchanged into
X-ray photons and redirected towards the disc by the corona or dissipated
by the corona according to the law
$$F_{cor} = A (r/R_C)^{-\beta}\eqno(2)$$
where A is determined by the available bolometric luminosity below $R_C$.
As fits of the model 2.3 (see Figure 3) to the data favoured the value
$\beta$ about 2, we fixed $\beta$ in the present model mostly at the value 2. 
Therefore our model has again only two independent parameters.

\subsection{Free--free emitting clouds}

As the existence of accretion discs in the centers of active nuclei is 
questioned (see e.g. Barvainis 1993) and optically thick clouds (e.g. Guilbert \& Rees 1988, 
Lightman \& White 1988, Sivron \& Tsuruta 1993) or optically thin clouds
(Antonucci \& Barvainis 1988, Ferland, Korista \& Peterson 1990) 
reprocessing hard X-ray radiation are suggested to be responsible 
for emission in the optical/UV/soft X-ray band we consider a cloud model 
as well. We assume that clouds are optically thin for absorption (nevertheless
they can be optically thick for scattering) and their radiation can
be well approximated as a bremsstrahlung emission at a single temperature
$T$. Such a spectrum has
two parameters: the temperature $T$ and normalization $V$
$$ F_{\nu} = V g_{ff}(\nu ,T) T^{-1/2}exp(-h\nu/kT) \eqno(3)$$
where the Gaunt factor $g_{ff}(\nu ,T)$ is taken from
Gronenschild \& Mewe 
(1978).  The normalization $V$ is proportional to the emitting volume and
to the square of the electron density.

We introduce this model due to its recent popularity but it is necessary
to realize that real cloud spectra would be dominated by strong emission 
lines, particularly for temperatures lower than $\sim 10^6 $ K
(e.g. Krolik \& Kriss 1995, Collin-Souffrin et al. 1996).

\subsection{Power law model}

In order to check how important is the correct determination of the spectrum
curvature in the successful models we also fit a simple power law to the same 
data,  with the slope and the normalization being free-parameters of this
formal model. Such a pure power law model has no direct physical 
interpretation although
a contribution of a power law component to the optical data of unspecified
origin has been broadly discussed (e.g. Loska \& Czerny 1990). Also, in the
course of analysis of the BLR contribution to the continuum an assumption was
made that the continuum can be approximated as a power law so the fit serves
as a test of this hypothesis.

\section {Method of analysis}
 
\subsection{Observational data}

\subsubsection{Observational campaign}

As the contribution of starlight might be considerable in the optical  band and
its influence strongly decreases with the studied frequency we concentrated
on the analysis of the continuum when the  UV data are available. 
The IUE observations of NGC 5548
covered the period December 1988 -- August 1989  and the fluxes at 1350\AA, 
1840\AA~, 2670\AA~ 
(SIPS) 
were taken from Clavel et al. (1991), and  5100 \AA~ from Peterson et al. 
(1992). We favoured the 5100 \AA~ value over the 4870 \AA~ flux from Peterson 
et al. (1991) as the second quantity is contaminated by the $H_{\beta}$ emission
and therefore less accurate, as argued by Peterson et al. (1992).
The optical measurements were carefully corrected by observers 
for the aperture effects
and reduced to a standard (A) set taken with the aperture 5.0''$\times$ 7.5''.

\subsubsection{Starlight problem}

The flux measured at 5100 \AA~ is thought to be contaminated by starlight 
from the host 
galaxy. The original optical data measured using different instruments and
apertures was corrected in such a way that only the starlight present in
the small aperture data $5.0'' \times 7.5'' $  remained. As the method did
not rely on any specific spectral shape of the starlight the adopted 
procedure was
actually independent from the nature of the extended emission. 

However, in order to model the intrinsic variability of the nucleus 
all the starlight 
contribution had to be removed from the data. For that purpose we assumed that
the shape of the starlight in NGC 5548 was well represented by the emission 
from the nucleus of M31, as shown by Wamsteker et al (1990). 
We adopted the relative starlight 
contribution at 5100 \AA~ in the standard aperture equal $3.4 \times
10^{-15}$ erg/s/cm$^2$ after Romanishin et al. (1995).
Having such normalization, we subtracted the starlight from all four frequency
points although at 2670 \AA~ and 1840 \AA~ and clearly at 1350 \AA~ 
its contribution is actually negligible
(never higher than two  per cent).

\subsubsection{Balmer continuum and blended FeII lines}

The flux at 2670 \AA~ is additionally 
 contaminated by contribution from the BLR. Peterson
et al (1991) estimate  this contribution as 20 \%  of the measured flux.
However, the variations of the nuclear emission are followed by variations
of the Balmer continuum and blended FeII emission with possibly smaller 
amplitude and a delay of order of a few days. Maoz et al. (1993) therefore
made much more detailed analysis of the BLR continuum. They showed that the
typical delay of the emission due to the blended Fe lines was of order of
10 days and they determined the
total luminosity of this component in the band 2160 \AA~-- 4130 \AA
~in each data set.  We therefore calculate the correction to the 2670 \AA~
flux as a mean flux in this spectral band and we adopt 20\% of this value as an
error, as suggested by the authors. There may be a contribution of the
BLR to the 1840 \AA ~ at a level up to 10 \% but this effect is smaller than
the correction at 2670 \AA ~ and not studied so carefully so we did not 
include it while correcting the data.

\subsection{Fitting procedure}

We fit the models to the data calculating the value of $\chi^2$ at every
time point separately, so 
$$\chi^2 = \sum_{i=1}^4 {(y_{obs}^i-y^i(a_1,a_2))^2 \over \sigma_i^2}\eqno(4)$$
where $y_{obs}^i$ is the observed value and $y^i(a_1,a_2)$ is calculated from
any of our two-parameter model. Errors $\sigma_i$ of the measurements at 
1350 and 1840 \AA~ are taken directly from the Table 2 (SIPS) of 
Clavel et al. (1991)
while errors at 2670 \AA~ were calculated assuming the errors of the subtracted
BLR continuum equal to 20\% of the determined flux and the error of subtracted
starlight at 5100 \AA~ was taken equal to the error of the total flux at minimum.

We search for the minimum with respect to the parameters $a_1$ and $a_2$ at
every time point using the procedure AMOEBA from Numerical
Recipes (Press et al. 1989). In order to accept or to
reject the model we calculate the global value of the  reduced 
$\chi^2$ for a given
model (i.e. we sum the contributions to $\chi^2$ from all four 
frequencies in all
time points and divide by the number of the degrees of freedom, including
these connected with the global parameters) 
while the fits at every time point give us the value of the model parameters
as functions of time.

\section{Results}

\subsection{Fits of theoretical models}

The last two models, namely the free-free emission (model 2.9) 
and a power law (model 2.10) do not
include any global model parameters apart from the two parameters varying
with time so their analysis is the simplest.

\beginfigure{1}
\epsfysize=80mm 
\epsfbox[30 170 530 670]{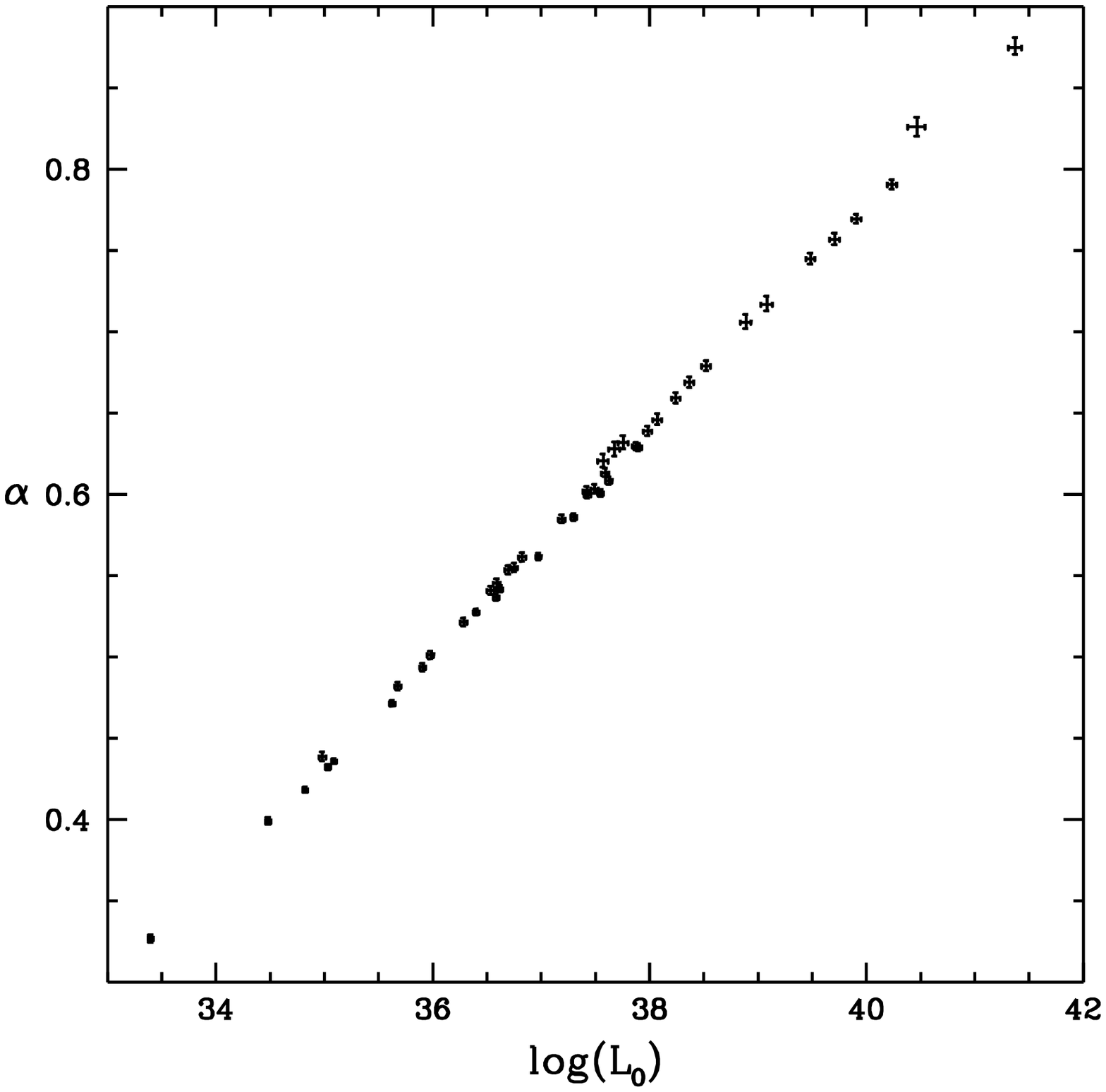}
\caption{{\bf Figure 1.} Energy index versus the normalization  constant
$L_o$  for a power law fit
(model 2.10). Error bars represent 90 \% confidence level.
}
\endfigure

The variations of the energy index with the luminosity clearly show the
well known trend in Seyfert spectra, i.e. the spectrum is harder when the
source is brighter. However, the curvature is the essential property of the
spectrum so the fit of the model 2.10 to the data is poor (see Table 1). 
Any model giving the $\chi^2 /dof$ above $\sim $ 1.52 can be rejected at the 
99.9\%
confidence level for the number of 
the degrees of freedom between 89 and 92 in presented models. It
means that the approximation of the continuum with a power law used to analyse
the BLR contribution is not very accurate. This may also mean that the 
continuum obtained by subtracting BLR contribution estimated on the basis of
a power law shape of the continuum may also contain some systematic errors.

\beginfigure{2}
\epsfysize=80mm 
\epsfbox[30 170 530 670]{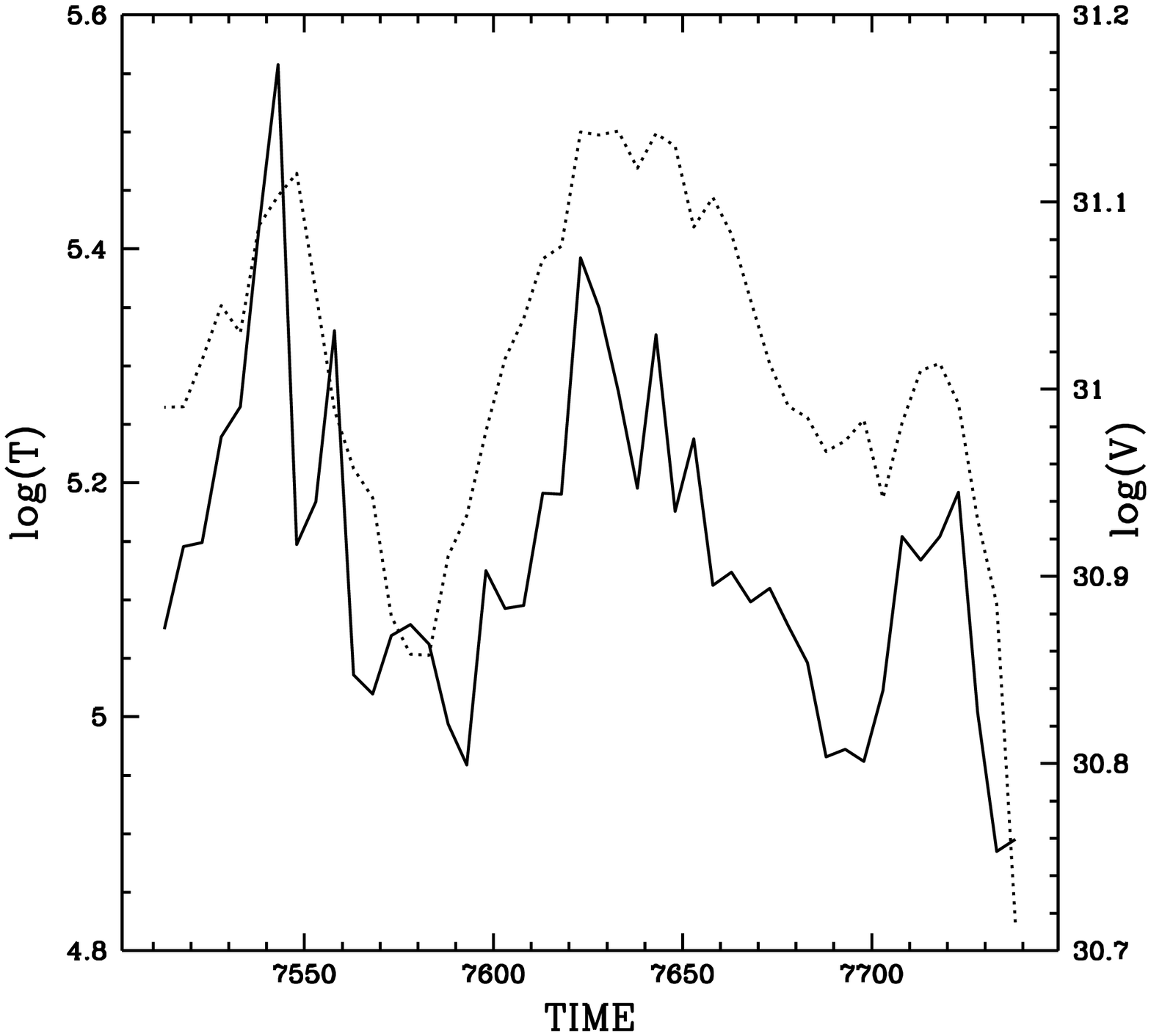}
\caption{{\bf Figure 2.} Temperature $T$ (continuous line) and emission 
measure $V$ 
(normalization
constant proportional to the square of the electron density and the gas 
volume; dotted line) as functions of time for a free-free emission (model 2.9)
}
\endfigure

A better fit is provided by the 
free-free isothermal model. The variable temperature
allows it to account for the variable spectral curvature. The total $\chi^2$ is
only marginally above 1 so the model is acceptable. The
model contains a 
strong correlation between the temperature and the normalization
factor. It means that either volume, or the density (or both) increase when the
temperature increases although the relative change in temperature is more than 
a factor 2 larger. This trend is not in contradiction with the expected behaviour.
If the timescale for expansion of the clouds under variable irradiation is long
we expect no variations of the volume measure. If this timescale is short and
the cloud adjusts itself to constant ionization parameter $\Xi$ to preserve
its thermal stability (e.g. Collin-Souffrin et al. 1996) the density may
indeed decrease with an increase of irradiation giving in effect an increase
in the measure of emission $V$ as defined in the model (i.e. proportional to
the product of the emitting volume and the square of the electron density). 
However, the presented result should not be treated as a satisfactory test
of the cloud model since the temperatures appropriate for the overall curvature
of the spectrum are not much higher than $10^5$K and the analytic free-free
formula is not an adequate description of the spectrum actually dominated by
line emission.

\begintable{1}
\caption{\bf Table 1. \rm The values of $\chi^2$ per one degree of freedom.
Models are numbered according to the corresponding paragraphs. Adopted values
of global parameters (mass of the black hole in solar masses 
and accretion rate in Eddington 
units or the index $\beta$), if appropriate, are given in column 2 and 3.
}

{\halign{%
\rm#\hfil&\hskip 23pt\rm#\hfil&\hskip 23pt\rm#\hfil&\hskip 23pt\rm\hfil#\cr
\noalign{\vskip 10pt}
\cr
Model&Mass/$M_{\odot}$&$\dot m$&$\chi^2$/dof\cr
\cr
2.1 &   $1 \times 10^7  $   &   0.01   &       1.05  \cr     
    &       ~~~~``          &   0.1   &        0.87  \cr     
    &       ~~~~``        &   0.5   &       0.99  \cr     
    &   $6 \times 10^7  $   &   0.01   &       0.87  \cr     
    &   $1 \times 10^8  $   &   0.005   &      0.93 \cr
    &   $5 \times 10^8  $   &   0.001   &       0.67  \cr     
2.2 &   $1 \times 10^7  $   &   0.01   &       1.03  \cr     
    &       ~~~~``           &   0.1   &       1.03  \cr     
    &       ~~~~``          &   0.5   &       1.03  \cr     
    &   $6 \times 10^7  $   &   0.1   &       1.01  \cr     
    &   $1 \times 10^8  $   &   0.005   &      1.02 \cr
    &   $5 \times 10^8  $   &   0.001   &       2.69  \cr     
2.3 &   $1 \times 10^7  $   &   0.01   &       2.66  \cr     
    &        ~~~~``          &   0.1   &       2.86  \cr     
    &        ~~~~``         &   0.5   &       4.22  \cr     
    &   $6 \times 10^7  $   &   0.001   &      2.16  \cr
    &         ~~~~``         &   0.005   &      2.22  \cr
    &      ~~~~``          &   0.01   &      2.31  \cr
    &   $1 \times 10^8  $   &   0.005   &      1.76 \cr         
    &   $5 \times 10^8  $   &   0.001   &       0.61  \cr     
2.9 &    ---                        &   ---     &       1.15  \cr     
2.10 &  ---                         &    ---    &        2.68 \cr

\cr
Model&Mass/$M_{\odot}$&$\beta$&$\chi^2$/dof\cr
 \cr
2.4 & $1 \times 10^7  $  &   ---  &    92.65 \cr
    & $6 \times 10^7  $  &   ---  &    ~0.81 \cr
    & $1 \times 10^8  $  &   ---  &    ~0.81 \cr
    & $5 \times 10^8  $  &   ---  &    ~7.33 \cr
2.5 & $1 \times 10^7  $  &   ---  &    ~5.78 \cr
    & $6 \times 10^7  $  &   ---  &    ~0.93 \cr
    & $1 \times 10^8  $  &   ---  &    ~0.87 \cr
    & $5 \times 10^8  $  &   ---  &    ~7.33 \cr
2.6 & $1 \times 10^7  $  &   ---  &    92.31 \cr
    & $6 \times 10^7  $  &   ---  &    ~0.75 \cr
    & $1 \times 10^8  $  &   ---  &    ~0.76 \cr
    & $5 \times 10^8  $  &   ---  &    ~7.33 \cr
2.7 & $1 \times 10^7  $  &   ---  &    ~5.35 \cr
    & $6 \times 10^7  $  &   ---  &    ~2.65 \cr
    & $1 \times 10^8  $  &   ---  &    ~1.54 \cr
    & $5 \times 10^8  $  &   ---  &    ~7.33 \cr
2.8 & $1 \times 10^7  $  &   2.0  &    ~7.73 \cr
    & $6 \times 10^7  $  &   1.8  &    ~5.34 \cr
    &    ~~~~``       &   2.0  &    ~0.94 \cr
    &      ~~~~``     &   2.2  &    ~2.47 \cr
    & $1 \times 10^8  $  &   1.8  &    ~0.55 \cr
    &    ~~~~``  &   2.0  &    ~0.51 \cr
    &  ~~~~``  &   2.2  &    ~0.48 \cr
    & $5 \times 10^8  $  &   2.0  &    ~7.33 \cr
\cr}}

\endtable

All the remaining models contain some global model parameters. 
We fix the value of the mass of the central black hole 
on the basis of available information. However, as
the adopted value can be questioned, we show the
major trends in the dependence on global parameters in Section 4.3. 
Other parameters ($\beta$ in  model 2.8 and $\dot m$ in models 2.1, 2.2 and
2.3) are chosen arbitrarily but the dependence of the fits on these 
parameters were tested.

The mass of the central massive black hole seems to be known relatively well.
Its value most probably lies between $5 \times 10^7
M_{\odot}$ and  $6 \times 10^7
M_{\odot}$ (Rokaki et al. 1993), and 
X-ray observations confirm the inclination
angle $\sim 20^o - 30^o$ used to derive these limits. 

X-ray reprocessing also
supports this value directly. 
The delay of the UV emission with respect to the X-ray data not more that
by two days indicate that the reprocessing region is of order of
$5 \times 10^{15}$ cm. The same region should be responsible for the formation
of the broad $K_{\alpha}$ Fe line. The width of the line larger than 
35 000 km/s corresponds to the Keplerian orbit below $\sim 36 r_g $. The
two dimensions are in mutual agreement for the mass of the central black
hole about $6 \times 10^7 M_{\odot}$. 

Direct modelling of the optical/UV data with an 
accretion disc gave the upper limit for the value of the black hole 
$5 \times 10^8 M_{\odot}$ (Loska et al. 1993).
The estimates of the mass of the black
hole based on the kinematics of the Broad Line Region lead to the value
about $10^8 M_{\odot}$ (e.g. Wanders et al. 1995) or somewhat smaller  
(Done \& Krolik 1996). 

Therefore in our basic analysis we adopt the value of the mass of the
black hole of $6 \times 10^7 M_{\odot}$, obtained from the most accurate
estimates and consistent with all
available  constraints.

For this value of the central black hole we found that
only the models 2.3 and 2.7 do not fit the data. All other models
well represent the behaviour of the optical/UV flux. Fits of model 2.3 
with larger and smaller values of $\dot m$ does not improve the fit
considerably as the model only weakly depends on $\dot m$. 
Model 2.7 does not include additional global parameters,
apart from the mass of the black hole.

This result is not surprising
as all the models were designed in such way as to have enough of flexibility.
However, the constraints on the model parameters lead to very interesting
results.

\beginfigure*{3}
\epsfxsize=180mm 
\epsfbox{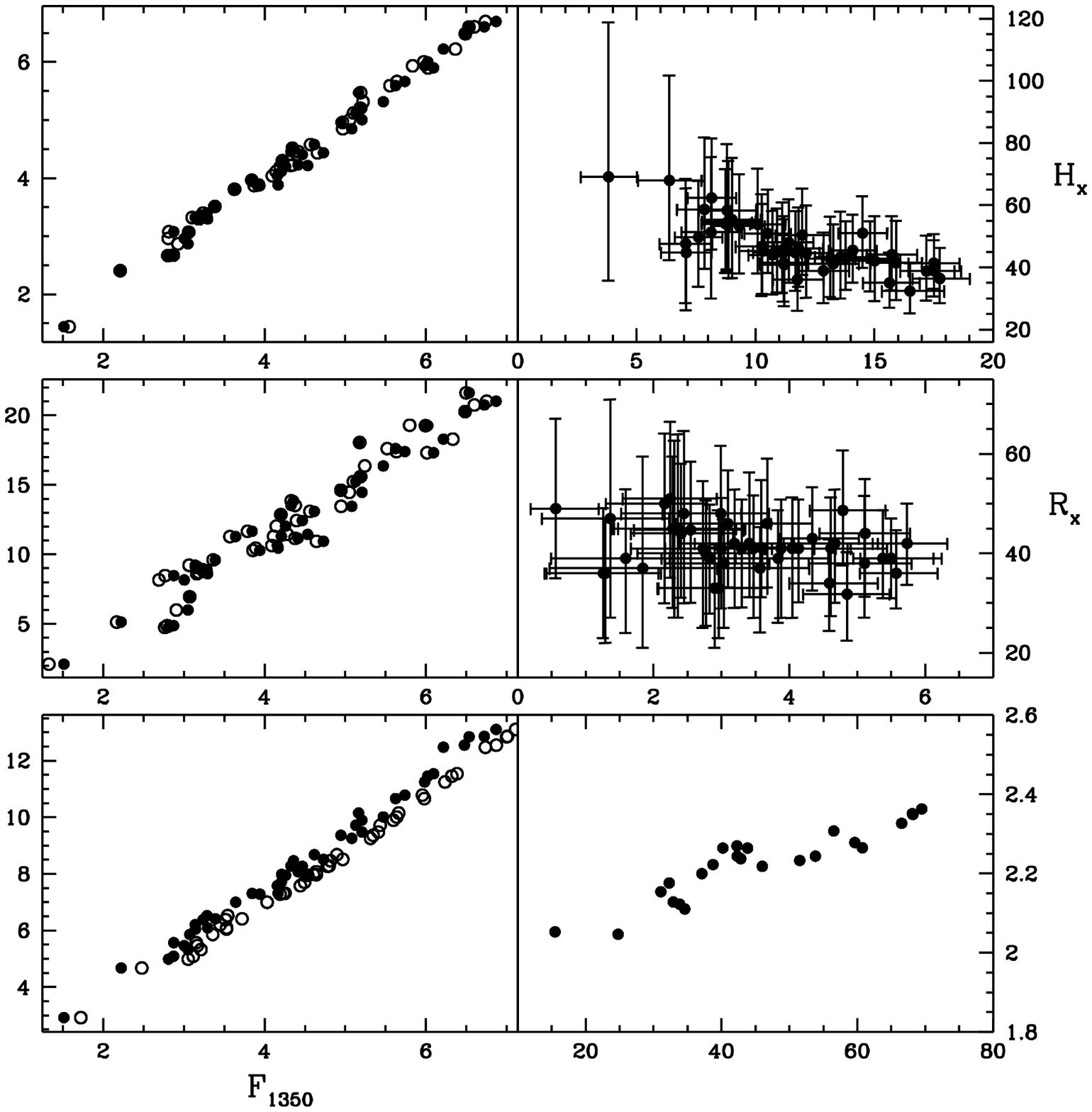}
\caption{{\bf Figure 3.} Stationary models: model 2.1 - upper panel,
model 2.2 - medium panel and model 2.3 - lower panel. The left side
of the diagram shows the dependence of the X-ray bolometric luminosity of the
model $L_X$ measured in $10^{44}$ erg/s versus the flux at 1350 \AA~ in
units $10^{-14}$ erg/s/cm$^2$/\AA ~ as measured directly (full dots) and
as derived from the best fit model (open circles). Right side of the diagram
shows the dependence of the height of the point-like source $H_X$ in $r_g$
units, the size of the optically thin part of the disc $R_X$ in $r_g$ and
index $\beta$, for models 2.1, 2.2, 2.3 correspondingly, as functions of the
X-ray bolometric luminosity, this time 
measured in units of the accretion disc viscous 
luminosity. Error bars represent 90\% confidence level.
 } 
\endfigure

Of the three stationary models, the first two are clearly acceptable which
means that we cannot differentiate between the point-like X-ray source and
an extended source of X-ray emission.  
However, it is interesting to note
that in both cases involved X-ray luminosity is considerably larger (factor
2 to 17) than
the luminosity viscously liberated in a disc. It is generally 
not surprising as the
variable irradiation has to account for the observed amplitude of the
flux changes. 

The second property the two models
have in common is the relatively large height of the X-ray source above the
disc plane; the radius of the optically thin sphere or the height of the
point-like source above the disc plane is of order of 30 - 60 $r_g$. Such
a constraint was by no means imposed on the models.

Among the non-stationary models, only model 2.7 is not acceptable.
The remaining models provide good fits to the data. Again, all of them
show very similar behaviour. The variability is driven by the change in
accretion rate (by a factor up to 3) around a value of order of 0.2. The
cut-off radius vary less  and in all cases it is 
anticorrelated with the accretion rate; the effect is the strongest in model 
2.7 and almost invisible in model 2.4.

One of these models (model 2.4) does not include any irradiation by X-rays. 
The model
fits the data well. For the value of the mass of the central black hole 
adopted in these calculations the cut-off  radius
is practically  uncorrelated with the accretion rate. Nevertheless, its 
presence is the essential
part of the model. Fits of the same model with the cut-off radius fixed
at the mean value ($\sim 16.5 r_g$) are much worse ($\chi^2 = 3.00/$dof). 
It might mean that
the transition from the optically thin to optically thick solution in the
innermost part of the disc is not determined by the accretion rate but by
some other external parameter, e.g. magnetic field. Although not intuitive,
such a possibility cannot be easily excluded.

If the irradiation is allowed the variations of accretion rate are of smaller
amplitude than in the model 2.4 as they are enhanced by absorbed X-rays.
Although in all models there is a weak anticorrelation of the cut-off radius
and the accretion rate, the liberated X-ray luminosity is directly correlated
with the accretion rate although the dependence is somewhat weaker than
linear. 

In these models the height or the size of the X-ray source is about a factor 2
smaller than in models with constant accretion rate so the match consistent 
with $K_{\alpha}$ constraints is much easier to achieve. 

The two models (2.3 and 2.7) are unable to fit the data as they do not produce
enough of the spectrum curvature in the UV band. They well represent the
NGC 5548 when it is bright but they are unable to model the spectrum when the
object is fainter and the maximum of the spectrum is clearly seen on the
log($\nu f_{\nu}$) vs. log$(\nu)$ diagram. However, this conclusion is
changed if we allow for larger mass of the central black hole (see Sect. 4.3). 

It is interesting to note that anomalously high continuum measurement at 
optical wavelength on JD 2,447,546 discussed by Peterson et al. (1991)
is reproduced by models 2.3, 2.7, 2.8, 2.9 and 2.10 but not by the others.

\subsection{Constraints from the X-ray observations}

There was no continuous coverage by any X-ray satellite during the IUE
campaign analysed in detail in this paper. However, shorter simultaneous
observations were carried out with Ginga satellite (Clavel et al. 1992). 
Also recent high quality  ASCA
observations put constraints on the models. These additional requirements
can be used to reduce the number of models allowed by the optical/UV data
alone.

\beginfigure*{4}
\epsfxsize=180mm 
\epsfbox{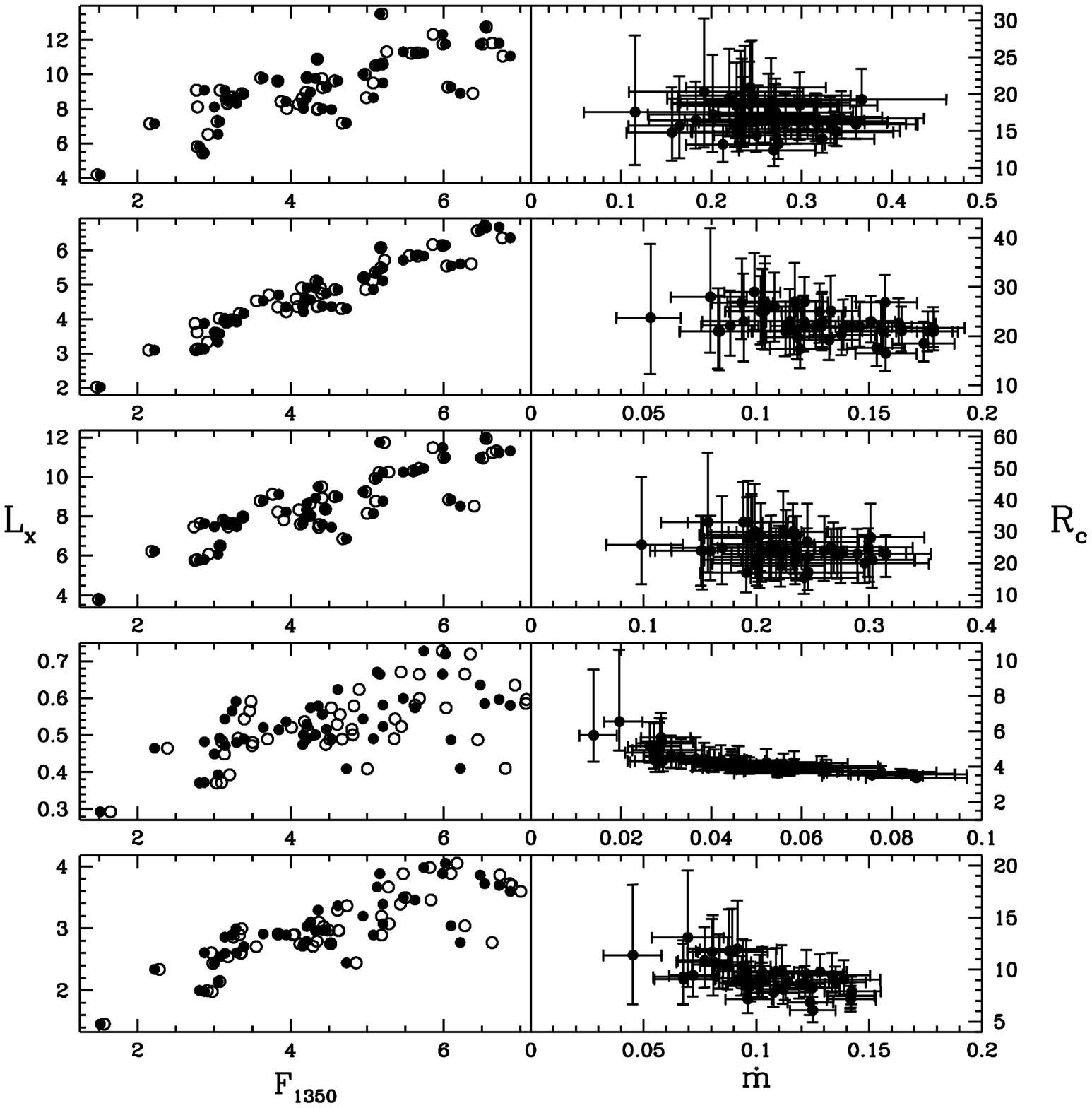}
\caption{{\bf Figure 4.} Non-stationary models: panels from the top to the
bottom show models 2.4, 2.5, 2.6, 2.7 and 2.8. The left side
of the diagram shows the dependence of the X-ray bolometric luminosity of the
model $L_X$ measured in $10^{44}$ erg/s versus the flux at 1350 \AA~ in
units $10^{-14}$ erg/s/cm$^2$/\AA ~ as measured directly (full dots) and
as derived from the best fit model (open circles). 
In the model 2.4 the X-ray energy shown is the 
total energy available below the cut-off radius; most of this energy is
advected and only an arbitrarily low fraction of it might eventually 
emerged in the form of X-rays. Right side od the diagram
shows the dependence of the cut-off radius  of the disc $R_C$ in $r_g$ 
 as function of the
accretion rate 
measured in Eddington units.  Error bars represent 90 \% confidence level.
 } 
\endfigure

The simplest constraint
comes from the mean bolometric X-ray luminosity.
Although its value is not known accurately due to the uncertainty of the
high energy extension of the hard X-ray power law but a reasonable estimate
gives value up to $5.5 \times 10^{44}$ erg/s (Done et al. 1995) when the
source is bright (for an adopted value of the mass of the central black hole
this luminosity translates into $\dot m = 0.08$). 
During the campaign analysed
in this paper the source was generally dimmer. All the models containing
the irradiation should return an X-ray luminosity of order or below this limit.
This condition is satisfied by models  2.1, 2.5 and 2.8 
(among the models with acceptable $\chi^2$) but not by models
 2.2 (for $\dot m=0.1$), and 2.6  (see 
Fig. 4 and Table 1). Allowing for lower 
$\dot m$ in model 2.2 still increases the luminosity. Larger value of 
accretion rate leads to acceptable luminosities but at the same time it 
increases the cut-off radii to unacceptable values larger than 36 $r_g$ 
(see below).

The constraint does not apply to model 2.4 as in this case the energy
flux available for X-ray production is mostly advected below the horizon
of the black hole.

\beginfigure{5}
\epsfysize=80mm 
\epsfbox[30 170 530 670]{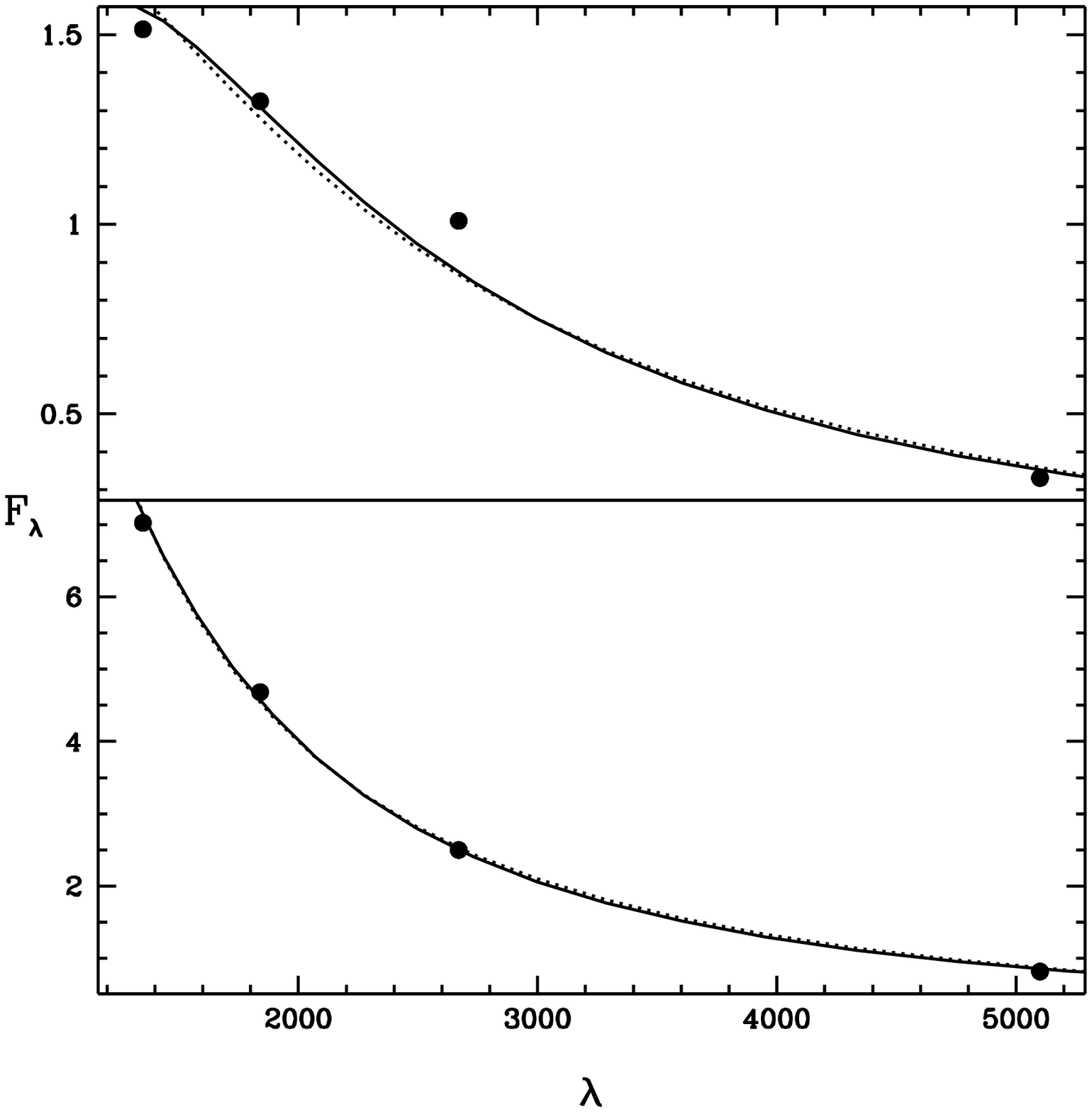}
\caption{{\bf Figure 5.} Observed optical/UV spectrum of the nucleus when
the source was faint (date 7738 - upper panel) and when it was bright 
(date 7622 - lower panel) are shown with dots. Best fits to the data by
the corona model 2.8 is marked as continuous line and the best fit of 
the free-free model 2.9 is marked with a dotted line. $F_{\lambda}$ is given 
in units $10^{-14}$ erg/s/cm$^2$/\AA~ and $\lambda$ in \AA.
}
\endfigure

The result of the X-ray monitoring was the proportionality of the X-ray
luminosity and 1350 \AA~ flux, with flattening when the 1350 \AA ~ flux
was above $5 \times 10^{-14}$ erg/s/cm$^{2}$/\AA. Therefore 
in Fig. 3 and Fig. 4  we plot the X-ray luminosity derived
from models versus 1350 \AA ~ flux. 
We see that the scatter on the diagrams showing stationary models is much
smaller than in the observational data in Fig. 4 of Clavel et al. (1992).
We can also use the formula from Clavel et al. (1992) to predict the 
range of the 2-10 keV flux expected from the observed range of the 1350 \AA ~
flux. If we additionally use the bolometric correction of Done et al. (1995) 
we estimate the total X-ray luminosity range to be $(1.4 - 5.1) 
\times 10^{44}$ erg/s. X-ray luminosities derived for the 
models 2.1, 2.5 and 2.8 cover just this range. 

The strength of the observed $K_{\alpha}$ line 
requires that approximately half of the X-ray flux
is reprocessed by the disc and the other half
 reaches an observer without interaction with the disc.
This requirement is approximately 
satisfied by all models in which irradiation is present
as the optically thin part of the disc is never significantly larger than
the height, or extension, of the X-ray source so almost half of the flux
can be intercepted.

ASCA observations in turn constrain the geometry of the models. According
to these data (Mushotzky et al. 1995) the $K_{\alpha}$ iron line has the 
FWHM larger than
35 000 which corresponds to Keplerian orbits below $\sim 36 r_g$. It has two
consequences.

The first limit imposed on the models is that the optically thick part of the
disc extends considerably below $36r_g$. This condition is satisfied by
models 2.1, 2.4, 2.5 and 2.8 as well as by some other models rejected on the
grounds of their high X-ray luminosity. 

The requirement that most of the line form at about $36 r_g$ and below put
also 
constraint on the location of the X-ray source. In the case of a point-like
source illuminating the plane most of the reprocessing 
takes place in a region with a radius about ${\sqrt 3}$ of the source 
height if the disc extends down to the black hole so this height should be of 
order of 25 $r_g$ or somewhat smaller. This requirement is not satisfied
by the stationary model 2.1 as the fitted X-ray source height in this model 
covers the range of 35 to 70 $r_g$.

In the case of a model with inner radius cut-off $R_C$ and the X-ray 
source placed at the height $R_C$ (model 2.5) half of the reprocessing
takes place below the radius ${\sqrt 7} R_C$ which limits $R_C$ to
values smaller than $\sim 14 r_g$.
This requirement is not satisfied  since values  as large as 
30 $r_g$ are reached. 

In model 2.4 we assumed that the energy released below the cut-off radius
is mostly advected under the horizon. However, the comparison of the observed
X-ray luminosity with the available one (see Fig. 4) shows that only $\sim
50$ \% of the energy should be advected and the remaining $\sim 50$ \% is 
necessary to explain the observed direct X-ray emission and the reflected
component. Such a considerable irradiation flux modifies the distribution
of the effective temperature, contrary to the initial assumption. Therefore the
model 2.4, although satisfactory from the point of view of the optical/UV data,
either does not produce X-rays (contrary to observations) or is not 
self-consistent (influence of X-rays cannot be neglected).

Model 2.8 is therefore the only disc model (for a mass of the black hole 
equal $6 \times 10^7 M_{\odot}$) fully in agreement with the 
available observations. The other acceptable description of the optical/UV
data gives  the free-free model (model 2.9); in this case the constraints 
from the X-ray observations are difficult to apply as it would require
the inclusion of additional parameters into the model representing the
geometry of the cloud distribution.

\subsection{Dependence on global parameters}

We explored the dependence on global parameters by fitting models for a few
values of the global parameters involved.

The dependence of the results on the assumed mass of the central black hole
is essential in all non-stationary disc models. Since we explicitly limited the
accretion rate to remain below the Eddington value all non-stationary models
give unacceptable fits if the mass of the black hole is $1 \times 
10^7 M_{\odot}$. If the black hole is very massive ($5 \times 10^8 M_{\odot}$)
the temperature of the disc becomes too low to account for the far UV emission
although models tend to settle themselves onto the solutions without 
effective cut-off
(i.e. $R_C=3r_g$) in order to ease this problem. 

The value of the mass of
the black hole about $1 \times 10^8 M_{\odot}$ are optimal for all 
non-stationary models, independent on geometry. The values of $\chi^2$ 
for all models are smaller or comparable to the values obtained for 
$6 \times 10^7 M_{\odot}$; even models 2.3 and 2.7 are acceptable. What is
more, the models ruled out on the basis of having too extended reprocessing
region (models 2.1 and 2.5) are perfectly consistent with X-ray data as
the cut-off radius decreases with an increase of the mass of the black hole.
Also the X-ray bolometric luminosity is smaller and the
predicted values agree within $\sim$ 50 \% (apart from model 2.7,
as before)
with expected values (see Section 4.2).
This trend is related to the drop of the disc temperature with an increase of
the mass of the central black hole and no need for a large cut-off radius
in order to explain the observed low temperature of the thermal emission. 
However, it is interesting to note that 
again the best model is model 2.8, this time purely on the basis of
having the lowest reduced $\chi^2$.  

As the stationary models have an energy source of unspecified nature we
did not introduce any apriori constraint on the luminosity. As a result,
most models fit the data for all values of the mass of the central black hole.
They depend very weakly on the accretion rate of an underlying accretion disc
as the X-ray luminosity strongly dominates disc luminosity. Models
for low value of the mass of the black hole marginally satisfy the 
constraints imposed by the total X-ray luminosity (the brightest models
exceed the estimated limit only by 40\%). Models for high mass are too bright
by a factor up to 4.
Therefore again the central mass at about $6 \times 10^7 M_{\odot}$ is 
strongly favoured.  

The dependence of stationary models on the adopted value of the accretion rate
of the disc is very weak as the X-ray luminosity strongly dominates and the
contribution of the radiation flux due to the disc viscosity is negligible in
comparison with the thermalized X-ray flux.

\subsection{Properties of the best models}

Constraints from the IUE observational campaign and from the available 
X-ray data reduced the acceptable models of the nucleus of the Seyfert galaxy
NGC 5548 to just two models: non-stationary accretion disc with a corona
(model 2.8) and free-free emitting clouds (model 2.9). We cannot favour any
of the two models as they are less specific than other models as for the geometry
and no additional constraints are available for them.

Both models fit the optical/UV data well as the actual spectra predicted by
the models in this frequency band are almost identical. We can see it in
Fig. 5 which shows the plot of the two model spectra in one of the highest
luminosity states (date 7622) and one of the lowest states (date 7738). The
high state spectrum is almost a power law while the low state has significant
curvature (actually, it shows a maximum in $\nu F_{\nu}$ diagram). Both models
follow these changes very well. Both models in low state are somewhat 
underluminous at 2670 \AA~ which probably indicates systematic error in the
subtraction of the BLR contribution.

Some of the properties of the model 2.8 are shown in Fig. 4. 
The accretion rate (measured in the Eddington units) vary between 0.05 and
0.14. The cut-off radius varies
between $6 r_g$ and $13 r_g$ 
and it is clearly anticorrelated with the X-ray 
bolometric luminosity. It means that the strength of the corona decreases with
increasing accretion rate. Such a trend is in agreement with a period of
even stronger
UV brightening (Clavel et al. 1992); anticorrelation of cut-off radius with 
accretion rate may eventually lead to saturation or even a decrease of the 
available X-ray flux when cut-off radius approaches $3 r_g$ (see Fig. 6).   

\beginfigure{6}
\epsfysize=80mm 
\epsfbox[30 170 530 670]{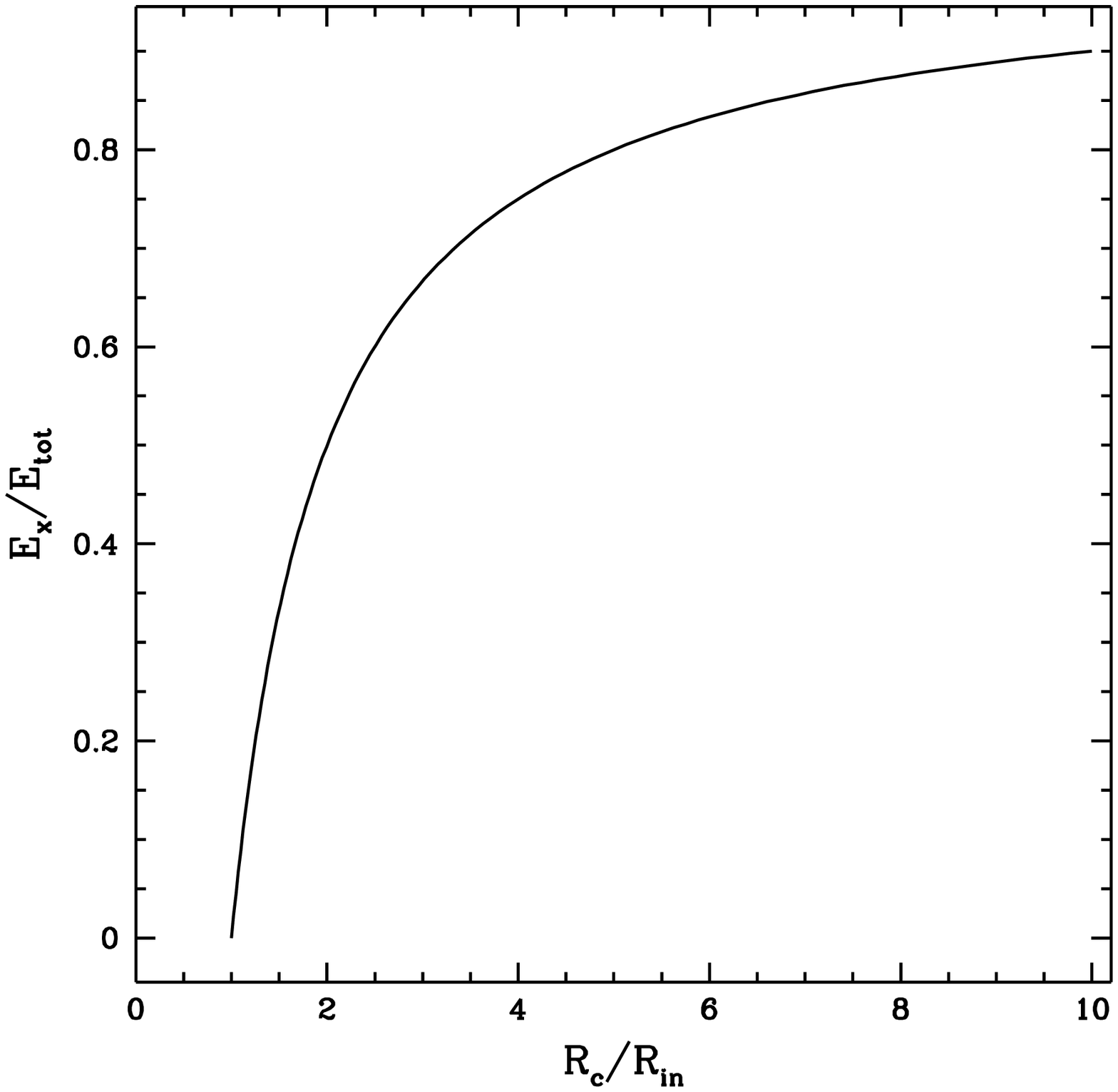}
\caption{{\bf Figure 6.} Fraction of gravitational energy available for the
X-ray emission as a function of the cut-off radius measured in units of the
radius of the marginally stable orbit equal $3 r_g$.
}
\endfigure

\section{Discussion}

As it was already concluded in the first paper on the monitoring campaign
of the Seyfert galaxy NGC 5548 (Clavel et al. 1991) 
the optical/UV emission is dominated 
most of the time by the
reprocessing of the X-ray radiation and further observations confirmed this
conclusion. Our models also support this trend.

The analysis presented in this paper favours one among the disc-type 
models of the nucleus:
disc/corona model.

We found also the cloud model 
to be acceptable and at present we are unable to differentiate between them.
The comparison is however unfair since 
acceptable disc models had to satisfy several very strong constraints based
by X-ray data whilst optically thin clouds were introduced as purely parametric
model without reliable description of their spectra and without 
a global scenario which would allow to use X-ray constraints to
this model as well.

The disk/corona model depends  on additional global parameters, including
the mass of the central black hole. We adopted the value of $6 \times 10^7
M_{\odot}$ (for arguments, see Section 4.1). Lower $\chi^2$ values for
all models, including disc/corona model, were obtained for somewhat
larger value ($ 1 \times 10^8 M_{\odot}$). However, an attempt to derive 
the value of the mass of the black hole directly from fits will return 
this value with a large error. 

Other disc type models were ruled out by the fact that models had to 
satisfy two opposite requirements: (i) the emission region had to have low
temperature in order to explain the roll-over of the spectrum in UV band when
the object is faint; for a fixed mass, it was achieved by an 
appropriately large
cut-off radius, (ii) the X-ray reflection component has to form close to the
black hole so the cut-off radius has to be small enough to allow the disc
to be present close to the black hole. As the disc temperature for a given
value of the gravitational radius drops with an increase of the mass of the
black hole those conflicting requirements are easy to satisfy when the
black hole is more massive. In any case the disc emission, particularly when
the nucleus is faint, does not extend far into EUV spectral region and is
not expected to be seen in soft X-rays.

As the disc/corona 
model is non-stationary, it poses a question whether any changes
of the accretion rate are possible in such a short timescale. In order to 
estimate that we have to realize that the change in the accretion rate at the
innermost part of the disc does not require the actual redistribution of the
mass in the disc but it may happen in a thermal timescale, as a transition
from one state to another. As the estimates of the timescales of disc/corona
system are not available we may use just the thermal timescales of the standard
disc based on the viscosity as described by Shakura \& Sunyaev (1973). These
timescales, as functions of the wavelength, 
were carefully estimated by Siemiginowska \& Czerny (1989). Assuming the
luminosity to the Eddington luminosity ratio as derived from the model (i.e.
0.05 - 0.14) and the observed range of the $F_{\lambda}$ at 1350 \AA ~ we
obtain the timescales below 1 day for the viscosity parameter $\alpha$
equal 0.1 which is well within the required limits. 

The second model representing the optically thin clouds is also acceptable. 
Unfortunately it is not specific enough in its present form to allow any
tests based on the properties of X-ray emission. This model may also have
difficulties to account for the temporary brightening in UV or in soft X-rays
which were sometimes observed when the source was bright (but not during
the campaign analysed in this paper). To account for such a phenomenon it
would be necessary to allow for some 
viscous dissipation within clouds. However,
no specific models of such a clumpy disc are available at present.

The similarity of the two model spectra when fitted to the optical/UV data
clearly shows that it is very difficult to differentiate between the two
models. Careful reduction of the data, removal of the starlight and the
BLR contribution are essential but not enough to favour any of the two
possibilities, and actually even more models are acceptable if the mass of
the central black hole is allowed to be somewhat higher.

Therefore, the problem lies clearly in too small dynamical range of the studied
spectra. However, the solution of the problem is not easy. 

It is relatively
simple to include an additional point at $\sim 8000$ \AA, as the entire
spectra are in principle available although their reduction is by no means
simple. 

Extension of the monitoring from optical to soft X-ray band would 
in principle help
significantly but it would make the problem of modelling very complex. 
Both optically thin clouds and disc/corona model spectra cannot be represented
by simple models considered in this paper (see Collin-Souffrin et al. 1996
for clouds and \. Zycki et al. 1994 for disc reprocessing). Reliable models
are not available yet for the soft X-ray band. However, as soon as acceptable
models are proposed such broader fits may be strongly constraining.

More direct use can be made from hard X-ray band monitoring if the data is
accurate enough to show the variations in the reflected component. 
The comparison
with such a data would not necessarily require much more sophisticated
models than currently described. Particularly interesting data would be 
collected when the source is rather faint since such observations might
determine whether the innermost part of the disc actually undergoes a
structural change or whether accretion rate is temporarily reduced. 
However, such monitoring is beyond the
present possibilities.

\section{Conclusions}

On the basis of 
the analysis of the IUE observational campaign of the Seyfert galaxy NGC 5548
and the constraints from the X-ray observations we show that the nuclear 
emission can be fitted by a model of a non-stationary
accretion disc with a corona. Simple analytic free-free formula 
representing the emission of a distribution of clouds optically thin for 
absorption gives an
acceptable description of the optical/UV data alone.
There is no possibility to distinguish between these two models on the basis
of the present data as it is not clear how to imply the 
constraints from the X-ray data in the case of cloud model. 
Any further progress can be only made if the research
is extended either towards careful studies, both observational and 
theoretical,  of the spectral features like emission lines and edges, or 
towards
broadening of the spectral range towards soft X-rays and/or hard
X-rays.


\section*{Acknowledgments}  We would like to thank Chris Done, the referee,
for extremely helpful remarks which lead to clarification of the line of
argument.
We are grateful to Andrzej So\l tan for very 
helpful discussions of the statistical problems, to Asia Kuraszkiewicz
for the discussions of the starlight subtraction and to Aneta Siemiginowska
for her subroutine to compute the Gaunt factor.
This project was partially supported by 
grants no.\ 2P30401004  and no. \ 2P03D00410 of the Polish State 
Committee for Scientific Research. 
This research has made use of the NASA/IPAC Extragalactic Database (NED)
operated by the Jet Propulsion Laboratory, Caltech.



\section*{References}

\beginrefs
\bibitem Antonucci, R.R.J., Barvainis, R., 1988, ApJ, 332, L13
\bibitem Barvainis, R., 1993, ApJ, 412, 513 
\bibitem Clavel, J. et al., 1991, ApJ, 366, 64
\bibitem Clavel, J. et al. 1992, ApJ, 393, 113
\bibitem Collin-Souffrin, S., 1991, A\&A, 249, 344
\bibitem Collin-Souffrin, S., Czerny, B., Dumont, A.-M., \. Zycki, P.T., 
         1996, A\& A (in press)
\bibitem Done, C., Krolik, J.H., 1996, ApJ (in press)
\bibitem Done, C., Pounds, K.A., Nandra, K., Fabian, A.C., 1995, MNRAS, 275,
         417
\bibitem Ferland, G., Korista, K.T., Peterson, B.M., 1990, ApJ, 363, L21
\bibitem Gronenschild, E.H.B.M.,  Mewe, R., 1978, A\&A Suppl.
         Ser., 32, 283
\bibitem Guilbert, P.W., Rees, M., 1988, MNRAS, 233, 475 
\bibitem Henri, G., Pelletier, G., 1991, ApJ, 383, L7
\bibitem Kotilainen, J.K, Ward, M.J., 1994, MNRAS, 266, 953
\bibitem Krolik, J.H., Horne, K., Kallman, T.R., Malkan, M.A., Edelson, 
         R.A., Kriss, G.A., 1991, ApJ, 371, 541
\bibitem Krolik, J.H., Kriss, G.A., 1995, ApJ, 447, 512
\bibitem Kusunose, M., Zdziarski, A.A., 1994, ApJ, 422, 737
\bibitem Liang, E.P., Li, H., 1995, A\&A, 298, L45
\bibitem Liang, E.P., Price, R.H.,  1977, ApJ, 218, 247
\bibitem Lightman, A.P., White, T.R., 1988, ApJ, 335, 57
\bibitem Loska, Z., Czerny, B., Szczerba, R., 1993, MNRAS, 262, L31
\bibitem Loska, Z., Czerny, B., 1990, MNRAS, 244, 43
\bibitem Malkan, M.A., 1991, in Structure and Properties of Accretion Disks,
         ed. C. Bertout et al., Paris, Editions Frontiers, p. 165
\bibitem Maoz, D., et al., 1993, ApJ., 404, 576 
\bibitem Marshall, H.L., Fruscione, A., Carone, T.E., 1995, ApJ, 439, 90
\bibitem Mushotzky, R.F., Fabian, A.C., Iwasawa, K., Kunieda, H., Matsuoka, M.,
         Nandra, K., Tanaka, Y., 1995, MNRAS, 272, L9
\bibitem Nandra, K. et al. 1991, MNRAS, 248, 760
\bibitem Nandra, K. et al. 1993, MNRAS, 260, 504
\bibitem Nandra, K., Pounds, K.A., 1994, MNRAS, 268, 405
\bibitem Narayan, R., Yi, I., 1995, ApJ, 444, 231
\bibitem Paczy\' nski, B.,  1978, Acta Astron., 28, 241
\bibitem Peterson, B.M., et al. 1991, ApJ, 368, 119
\bibitem Peterson, B.M. et al. 1992, ApJ, 392, 470
\bibitem Peterson, B.M., et al. 1994, ApJ, 425, 622
\bibitem Press, W.H., Flannery, B.P., Teukolsky, S.A., Wetterling, W.T., 1989,
         Numerical Recipes in Pascal: The Art of Scientific Computing, 
         Cambridge University Press, New York
\bibitem Rokaki, E., Collin-Souffrin, S., Magnan, C., 1993, A\&A, 272, 8
\bibitem Romanishin, W., et al. 1995, ApJ., 455, 516
\bibitem Ross, R.R., Fabian, A.C., 1993, MNRAS, 261, 74 
\bibitem Shakura, N.I., Sunyaev, R.A., 1973, A\&A, 24, 337
\bibitem Shapiro, S.L., Lightman, A.P., Eardley, D.M., 1976, ApJ, 204, 187
\bibitem Siemiginowska, A., Czerny, B., 1989, MNRAS, 239, 289
\bibitem Sivron, R., Tsuruta, S., 1993, ApJ, 402, 420
\bibitem Wamsteker, W., et al., 1990, ApJ., 354, 446 
\bibitem Wandel, A., Liang, E.P., 1991, ApJ, 283, 842
\bibitem Wanders, I. et al., 1995, ApJ, Nov.10 
\bibitem Ward, M., Elvis, M., Fabbiano, G., Carleton, N.P.,
         Willner, S.P., Lawrence, A., 1987, ApJ, 315, 74
\bibitem \. Zycki, P.T., Collin-Souffrin, S., Czerny, B., 1995, MNRAS, 277, 70
\bibitem \. Zycki, P.T., Czerny, B., 1994, MNRAS, 266, 653
\bibitem \. Zycki, P.T., Krolik, J.H., Zdziarski, A.A., Kallman, T., 1994,
         ApJ, 437, 597
\endrefs
\bye